\documentclass[sigplan,screen]{acmart}
\newcommand{\isuaffiliation}{%
  \affiliation{%
    \institution{Iowa State University}
    \city{Ames}
    \state{Iowa}
    \country{USA}
  }%
}

\usepackage{subfiles}
\usepackage{pgfplotstable}
\usepackage{booktabs}
\usepackage{adjustbox}
\usepackage{listings}
\usepackage{algorithm}
\usepackage{algpseudocode}
\usepackage{float}
\usepackage{tabularx}
\usepackage{array}

\renewcommand\footnotetextcopyrightpermission[1]{}

\AtBeginDocument{%
  }

\begin{document}

\title[MLIRBench]{MLIRBench - A Benchmark and Dataset for Compiler and Hardware-Aware Reasoning}

\author{Pranava Sai Maganti}
\authornote{These authors contributed equally to this work.}
\email{pranava7@iastate.edu}
\orcid{0009-0004-7249-6057}
\isuaffiliation

\author{Meghasyam Peddireddy}
\authornotemark[1]
\email{syam@iastate.edu}
\orcid{0009-0006-1968-3560}
\isuaffiliation

\author{Muzakker Hossain}
\email{muzakker@iastate.edu}
\orcid{0000-0003-2874-1380}
\isuaffiliation

\author{Kaining Gao}
\authornotemark[1]
\email{kaigao@iastate.edu}
\orcid{0009-0002-5580-8640}
\isuaffiliation

\author{Farhan Ahmed}
\email{fahmed@iastate.edu}
\orcid{0009-0008-7022-1235}
\isuaffiliation

\author{Ali Jannesari}
\email{jannesar@iastate.edu}
\orcid{0000-0001-8672-5317}
\isuaffiliation

\renewcommand{\shortauthors}{Maganti et al.}


\begin{abstract}

In recent years, large language models (LLMs) have become increasingly popular in software engineering and have performed well on different programming tasks. However, it is still difficult to know whether these models truly understand program behavior or whether they mainly rely on patterns learned from their training data. This distinction is particularly important for intermediate representations (IRs), where operations, dataflow, and execution semantics are explicitly represented. In this paper, we introduce \textbf{MLIRBench}, a benchmark for evaluating how LLMs reason about MLIR programs. MLIRBench uses a multi-agent pipeline to generate program mutations, validate their behavior through execution, and construct reasoning tasks from \texttt{torch} and \texttt{linalg} program pairs. The dataset contains \textbf{1,768} validated pairs, including \textbf{888} \texttt{linalg} pairs and \textbf{880} \texttt{torch} pairs, labeled as \textsc{EQUIV}, \textsc{DIVERGE}, or \textsc{ERROR}. We evaluated nine models, including five open-weight and four proprietary models. Gemini 3.1 Pro achieved the highest hard-label accuracy of \textbf{92.65\%} across all 1,768 pairs, counting unsuccessful predictions as incorrect. Across its
1,768 scored predictions, it achieved an aggregate score of \textbf{74.18} and a judge-assigned result-behavior score of \textbf{5.04}/10. These results show that strong semantic classification does not necessarily imply accurate prediction
of detailed execution behavior, highlighting a gap in current LLM reasoning over compiler IRs. By exposing this gap through execution-validated program pairs, MLIRBench provides a reproducible foundation for studying semantic reasoning over compiler IRs and related tasks such as program repair and semantic classification.

\end{abstract}



\settopmatter{printacmref=false}

\maketitle
\raggedbottom

\makeatletter
\fancyhead[LE]{\ACM@linecountL}
\fancyhead[RO]{\ACM@linecountR}
\makeatother

\setlength{\textfloatsep}{6pt plus 1pt minus 1pt}
\setlength{\floatsep}{5pt plus 1pt minus 1pt}
\setlength{\intextsep}{5pt plus 1pt minus 1pt}
\setlength{\dbltextfloatsep}{6pt plus 1pt minus 1pt}
\setlength{\dblfloatsep}{5pt plus 1pt minus 1pt}

\section{Introduction}

{
\setlength{\dbltextfloatsep}{6pt plus 1pt minus 1pt}

\begin{figure*}[t]
    \centering
    \includegraphics[
        width=1.0\textwidth,
        trim=10 8 10 8,
    ]{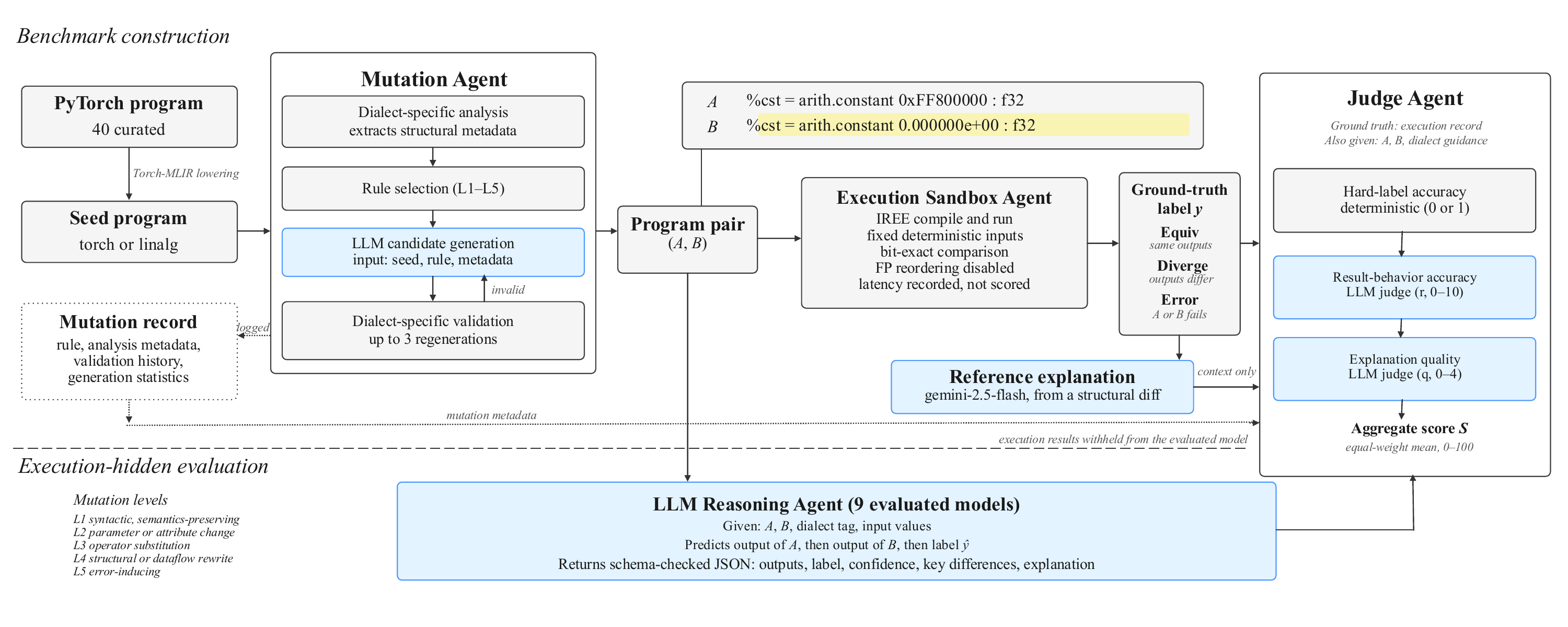}
    \caption{Overview of MLIRBench. \textbf{Benchmark construction (upper):} PyTorch programs are lowered to \texttt{torch} and \texttt{linalg} seeds. The Mutation Agent transforms a seed $A$ into $B$, guided by dialect-specific structural analysis, and records the mutation metadata. The Execution Sandbox Agent executes $A$ and $B$ on fixed inputs to determine the ground-truth label $y$. \textbf{Execution-hidden evaluation (lower):} each evaluated model receives only $A$, $B$, the dialect tag, and input values. The Judge Agent scores the model response using the execution record, mutation metadata, and reference explanation; only execution determines $y$. Blue boxes denote LLM calls. The code example shows a row-wise max reduction whose initial value is changed from $-\infty$ (\texttt{0xFF800000} in MLIR) to $0.0$.}
    \Description[Two-band MLIRBench pipeline from benchmark construction to execution-hidden evaluation]{The upper band constructs the benchmark. A PyTorch program is lowered to a \texttt{torch} or \texttt{linalg} seed. The Mutation Agent uses dialect-specific structural analysis, mutation-rule selection, LLM candidate generation, and validation to transform seed $A$ into program $B$ and produce a mutation record. The Execution Sandbox Agent executes $A$ and $B$ with IREE on fixed inputs and assigns the ground-truth label as Equiv, Diverge, or Error. The lower band evaluates models without execution access. The LLM Reasoning Agent receives $A$, $B$, the dialect tag, and input values, predicts the program outputs and label, and returns a structured response. The Judge Agent evaluates the response using the execution record, mutation metadata, and reference explanation, producing scores for label accuracy, result behavior, and explanation quality.}
    \label{fig}
\end{figure*}
}

LLMs have recently shown strong performance in many software engineering tasks such as code generation, bug fixing, test generation, and program understanding. Most existing studies and benchmarks evaluate these models on source-level programming languages like Python, C++, or Java. In these settings, models can often use natural language context and patterns learned from large amounts of training data. However, much less is known about how well LLMs can understand and reason about compiler intermediate representations (IRs), which describe programs at a lower level and expose more of the underlying program behavior.

Compiler IRs present a fundamentally different reasoning challenge from source code. Rather than expressing high-level programming abstractions, they expose the representations that compilers use to analyze, transform, and optimize programs. Reasoning about IRs often requires understanding dataflow dependencies, tensor shapes, type systems, memory movement, and dialect-specific semantics. Consequently, strong performance on source-code tasks does not necessarily translate to compiler reasoning. A model may generate correct code yet still struggle to determine whether two IR programs are semantically equivalent, whether a transformation preserves behavior, or whether a program should fail during verification, compilation, or execution.

The MLIR framework has become an important compiler infrastructure in areas such as machine learning, domain-specific languages, heterogeneous computing, and high-\\performance computing. MLIR is particularly interesting for evaluating LLMs because it combines higher-level program structure with lower-level compiler details. Through its different dialects, MLIR can represent tensor computations, linear algebra operations, hardware-specific transformations, and multiple stages of compiler lowering. As MLIR continues to gain adoption, it is useful to understand whether LLMs can assist with compiler-related tasks such as debugging, optimization, transformation validation, and compiler development.

Current benchmarks do not fully evaluate these capabilities. Many existing code benchmarks focus on code generation quality, compilation success, or passing unit tests. While these metrics are useful, they do not directly measure whether a model understands the semantic impact of compiler transformations. In MLIR, two programs may look very different while still producing the same output. On the other hand, a small modification to an operation, attribute, or indexing map can completely change program behavior. Some changes may even create invalid programs that fail verification, compilation, or execution. Therefore, evaluating semantic reasoning in compiler IRs requires a benchmark that is grounded in actual execution behavior.

To address this challenge, we introduce \textbf{MLIRBench}, an execution-grounded benchmark designed to evaluate semantic reasoning over MLIR programs. Figure~\ref{fig} shows the overall framework. MLIRBench uses a multi-agent pipeline to generate controlled mutations of MLIR programs, validate them through compilation and execution, and evaluate whether language models can correctly reason about their behavior. The Mutation Agent creates modified versions of programs at different mutation levels. The Execution Sandbox Agent verifies, compiles, executes, and tests these programs to establish ground-truth behavior. The LLM Reasoning Agent receives only the original and mutated MLIR programs and predicts their relationship without seeing execution results. Finally, the Judge Agent compares model predictions with the execution-grounded labels and computes benchmark scores.

MLIRBench focuses on three semantic outcomes: \textsc{EQUIV}, \textsc{DIVERGE}, and \textsc{ERROR}. An \textsc{EQUIV} pair contains programs that behave the same even if their structure differs. A \\ \textsc{DIVERGE} pair contains programs that compile and execute successfully but produce different results. An \textsc{ERROR} pair contains programs that fail during verification, compilation, or execution. Although these labels are simple, correctly identifying them often requires deep reasoning about compiler semantics and program behavior.

The benchmark includes multiple mutation levels and MLIR dialects. These mutation levels range from simple syntax-preserving changes to more challenging modifications such as parameter updates, operator substitutions, structural transformations, and error-inducing mutations. This setup allows us to analyze not only overall model accuracy but also the types of reasoning that are most difficult. For example, a model may correctly handle variable renaming but struggle when tensor shapes, affine mappings, or operation semantics are changed. Because of this, MLIRBench can serve both as an evaluation benchmark and as a source of training data for compiler-aware language models.

Our work is guided by five research questions. \textbf{RQ1:} Can large language models accurately determine semantic equivalence and behavioral differences directly from MLIR programs without execution feedback? \textbf{RQ2:} How does model performance vary across different mutation levels, from simple edits to complex structural changes and failures? \textbf{RQ3:} What reasoning mistakes do models make when evaluated using execution-grounded labels, and how do these differ from results reported by traditional code benchmarks? \textbf{RQ4:} Can execution-grounded supervision from MLIRBench
generalize to unseen program families across MLIR dialects?   {} \textbf{RQ5:} Can MLIRBench support compiler-grounded optimization-benefit prediction, and how does IR-level benefit relate to downstream RISC-V code generation?

Our contributions are:

\begin{itemize}
    \item We introduce \textbf{MLIRBench}, a benchmark for evaluating semantic reasoning over MLIR programs using execution-grounded labels.

    \item We develop a multi-agent framework that automatically generates, validates, executes, and labels MLIR program mutations across different reasoning levels.

    \item We create large-scale datasets of MLIR program pairs with execution-validated labels for equivalence, divergence, and error behavior.

    \item We evaluate both frontier and open-source language models on MLIR reasoning tasks and analyze their strengths and weaknesses across mutation categories and dialects.
    
    \item We further explore MLIRBench as a source of compiler-grounded supervision
for optimization-benefit prediction and semantic classification, including
family-disjoint evaluation across \texttt{linalg} and \texttt{torch} and

RISC-V backend analysis.
\end{itemize}

\section{Background \& Related Work}

\subsection{MLIR and Compiler Intermediate Representations}

Compiler intermediate representations differ from source-level programs in what they make explicit. Most source code leaves dataflow, types, memory layout, and iteration structure largely implicit, to be recovered by the front end; an IR fixes these properties as first-class structure so that later passes can analyze and transform programs soundly. Reasoning about an IR therefore means reasoning about the representation the compiler actually operates on, rather than about surface syntax.

MLIR is a multi-level compiler infrastructure organized around extensible dialects, allowing a program to be represented at several abstraction levels and lowered progressively from framework-level operators toward backend-specific code~\cite{9370308}. MLIRBench works with two of these dialects, \texttt{torch} and \texttt{linalg}, which together capture both framework-level operators and structured tensor computation; they are a subset of the MLIR ecosystem rather than its full extent. Within this design, a program exposes SSA dataflow, tensor shapes and element types, indexing structure, iterator semantics, and dialect-specific verifier constraints that determine whether a construct is well formed. These properties are key ones a semantic analysis must respect, and they are explicit in the IR rather than inferred. This makes MLIR a useful target for evaluating semantic reasoning: whether two MLIR programs behave identically generally requires following the structure the dialects encode, which is difficult to approximate from textual pattern matching alone.

\subsection{Execution-Grounded Compiler Testing and Program Equivalence}

A long line of compiler validation work treats observed program behavior as the reference for correctness. Differential testing compares outputs across programs or compilation settings and treats any observable difference as evidence of a semantic change~\cite{yang2011csmith,le2014emi}. The same methodology has been extended to MLIR through structured IR generation and fuzzing~\cite{wang2023mlirsmith,suo2024mlirod} and to deep-learning compilers through model-level test generation~\cite{liu2023nnsmith}. Grounding evaluation in execution is stronger than relying on annotations produced by a model or a heuristic, because the label reflects what the compiled program actually computes rather than what an annotator believes it should compute.

Cast in these terms, comparing two programs under identical inputs yields three outcomes of interest: behavioral equivalence, behavioral divergence, and compilation or runtime failure. MLIRBench uses exactly this distinction, labeling each program pair \textsc{Equiv}, \textsc{Diverge}, or \textsc{Error} from execution rather than from any model judgment. One subtlety is that not all behaviorally significant differences are visible by inspection: floating-point arithmetic is not associative, so a reordering that is algebraically neutral can still change the computed result~\cite{goldberg1991floating}. Because the labels are derived from bit-level output comparison under a fixed configuration, such differences are captured rather than assumed away.

\subsection{LLMs for Code Reasoning and Compiler IR}

Most existing benchmarks for code-capable language models operate at the source level. They measure program synthesis from natural-language specifications~\cite{chen2021evaluating}, probe code understanding through input--output and execution prediction~\cite{gu2024cruxeval,liu2024codemind}, evaluate under contamination-resistant conditions~\cite{jain2025livecodebench}, extend to real-world repository tasks~\cite{roy2026codesense,jimenez2024swebench}, and test semantic equivalence across program variants~\cite{wei2025equibench}. A related line measures whether models can produce performant kernels rather than merely correct ones~\cite{ouyang2025kernelbench}. These efforts characterize source-level and performance-oriented capability, but they do not directly target multi-level compiler IRs such as MLIR.

A smaller body of work targets compiler IRs directly. Models have been trained and evaluated on low-level IR and assembly for optimization and related tasks~\cite{cummins2025llmcompiler}, and studies of reasoning over single-level IR report that models parse IR syntax competently while struggling with instruction-level execution reasoning~\cite{jiang2025iir}. Reasoning over MLIR is distinct from both source-level reasoning and single-level IR reasoning: a model must track SSA dependencies, shape and element-type propagation, indexing and iterator structure, and dialect-specific legality, and it must do so across multiple levels of abstraction. We do not claim that any single prior result settles whether models possess such reasoning ability; the consistent observation is that compiler-internal reasoning is a separate and harder capability than source-level code competence.

\subsection{Benchmarking Compiler-Aware LLM Reasoning}

Relative to these efforts, MLIRBench occupies a specific position. Source-level reasoning and synthesis benchmarks~\cite{chen2021evaluating,gu2024cruxeval,jain2025livecodebench,liu2024codemind,roy2026codesense,jimenez2024swebench} do not operate on compiler-internal representations; equivalence and kernel benchmarks~\cite{wei2025equibench,ouyang2025kernelbench} evaluate source-level program equivalence and performance-oriented kernel generation, but do not target MLIR dialect semantics or MLIR verifier behavior; compiler testing and fuzzing~\cite{yang2011csmith,le2014emi,wang2023mlirsmith,suo2024mlirod,liu2023nnsmith} generate programs or mutations to test compiler correctness, but do not evaluate whether language models can reason about the resulting IR semantics; and compiler-IR language-model work~\cite{cummins2025llmcompiler,jiang2025iir} focuses primarily on LLVM IR or assembly rather than MLIR's multi-level, dialect-based structure. The gap these leave open is the direct evaluation of execution-hidden semantic reasoning over MLIR program pairs: a setting in which the evaluated model sees only the program pair, the dialect tag, and deterministic inputs when available, while executed outputs and execution-derived labels are withheld and used only for scoring. LLM-as-judge methods such as G-Eval evaluate natural-language outputs using an LLM-based scoring process~\cite{liu2023geval}, while ICE-Score applies LLM-based evaluation to code-generation outputs~\cite{zhuo2024ice}. However, neither directly measures execution-grounded MLIR semantics, dialect-specific legality, or compiler/runtime failure behavior, so MLIRBench derives hard labels from compilation, execution, and output comparison.

MLIRBench targets this gap, and it does so with a measurement that holds distinct abilities apart. Rather than reporting a single accuracy figure, it separates hard-label accuracy-whether a model assigns the correct semantic relationship-from result-behavior accuracy, which checks the concrete output behavior a model predicts, and from explanation quality. This separation surfaces cases where a model selects the correct label without correctly tracing program behavior, a distinction that matters directly for compiler-internal analysis and for tools that would depend on such reasoning.

\section{MLIRBench Framework}

\subsection{Framework Overview}

MLIRBench is a multi-agent benchmarking framework designed to evaluate semantic reasoning over MLIR programs using execution-grounded evaluation. Figure~\ref{fig} presents the complete workflow.

Starting from an input MLIR program and a target mutation level, the Mutation Agent generates controlled program variations. The Execution Sandbox Agent validates, compiles, and executes both programs to establish execution-derived ground truth. The LLM Reasoning Agent receives only the original and mutated programs and predicts their behavior without access to execution artifacts. Finally, the Judge Agent compares model predictions against the execution-derived ground truth and produces the final benchmark scores.

This design separates benchmark generation, execution, reasoning, and evaluation, which enables reproducible and execution-grounded assessment of compiler-oriented reasoning capabilities.

\subsection{Mutation Agent}

The mutation agent is responsible for generating controlled variations of an input MLIR program. Its primary objective is to produce program pairs that require varying levels of semantic reasoning, while maintaining a well-defined relationship between the original and the mutated program.

Given an input program, a dialect identifier, and a mutation level, the agent first performs a dialect-specific analysis. Separate analysis pipelines are implemented for the \texttt{linalg} and \texttt{torch} dialects to extract structural metadata, which is used to guide mutation generation and reduce invalid transformations.

MLIRBench organizes mutations into five increasingly challenging levels:
\textbf{L1:} semantic-preserving syntactic transformations;
\textbf{L2:} modifications of parameters and attributes;
\textbf{L3:} operator substitutions;
\textbf{L4:} structural and data-flow transformations;
\textbf{L5:} error-inducing mutations that violate compiler constraints and trigger verification, compilation, or runtime errors.

For each dialect and mutation level, MLIRBench maintains a collection of mutation requirements. These requirements describe the target transformation as well as the constraints that must be adhered to during generation. Examples include the renaming of SSA variables, operand substitution, operation swapping, structural rewriting, and mutations specifically designed to violate verification rules. The selected mutation requirement, along with the original MLIR program and the metadata obtained during analysis, is provided as input to a large language model to generate a candidate program for the mutation.

However, since LLM-generated IR code is frequently invalid, each candidate mutation undergoes automated validation using dialect-specific verification tools. Invalid mutations are discarded and regenerated. Up to three regeneration attempts are permitted for each mutation. Only mutations that successfully pass all required validation steps are accepted and forwarded to the Execution Sandbox Agent.

In addition to the original and the mutated program, the agent logs the selected mutation requirement, the analysis metadata, the validation history, and generation statistics. These artifacts provide complete and traceable documentation of the benchmark creation process and enable downstream analyses of mutation quality and generation behavior.

Through this analysis-driven and validation-based workflow, MLIRBench can generate large collections of targeted MLIR program pairs while ensuring syntactic correctness, dialect-specific consistency, and reproducibility. Additional implementation details for the Mutation Agent, including mutation inputs, prompt construction, validation, stored records, and common mutation failures, are provided in Appendix B.

\subsection{Execution Sandbox Agent}
\label{sec:execution_agent}
 
The Execution Sandbox Agent assigns the ground-truth divergence label for every
program pair in MLIRBench. For a pair $(A,B)$ it compiles and runs both programs
under identical deterministic conditions, compares their outputs, and records
whether they diverge. The label is computed from execution alone and never from
a language model, so the semantic ground truth is independent of the systems the
benchmark evaluates. Comparing the outputs of two compiled programs follows the
differential-testing methodology long used for compiler validation%
~\cite{yang2011csmith,le2014emi}, 
applied here at the level of MLIR program pairs.

We compile and run both programs with IREE~\cite{iree2023}. A pair is labeled success only when both programs compile and produce output; the remaining status classes are defined in Appendix C.4.
 
We compare the outputs of $A$ and $B$ with \path{numpy.array_equal}, which
requires bit-exact equality and applies no tolerance. Strict equality is
defensible here because we remove the compiler's sources of floating-point
reassociation rather than absorbing them into a tolerance. We disable the LLVM
optimization passes that reorder floating-point operations and compile $A$ and $B$ with the same flags,
so any surviving output difference is attributable to the mutation and not to the
backend. Floating-point addition and multiplication are not
associative~\cite{goldberg1991floating}, so an operation reordering that is exact
in $\mathbb{R}$ can still change the machine result. We treat such cases as
divergent, because the benchmark measures the behavior a model would observe from
real execution rather than from idealized arithmetic.
 
Both dialects run on deterministic inputs, constructed differently for each.
Linalg programs embed their inputs as \texttt{arith.constant} operations and
expose a zero-argument entry point; the embedded values come from the constants
already present in the program, with fixed fallback patterns where a tensor is
otherwise unconstrained. Torch programs cannot embed their inputs, since doing so triggers a compiler crash described in Appendix C.2, so they receive inputs at runtime. For each torch program we generate inputs that match its parsed input
signature, the shapes and element types it declares, through a fixed generation
procedure that is reproducible across runs. Because both mechanisms are
deterministic, two programs that declare the same input tensor types receive the
same inputs, and re-running a pair reproduces the same outputs and the same label.
Program $A$ and its mutation $B$ therefore receive identical inputs whenever the
mutation preserves the input tensor types. The exceptions are 25 torch T2 mutations that change those types; these cases are discussed in Appendix C.4 and Appendix C.5. Each pair is evaluated on one
fixed input set, so a mutation that changes behavior only on other inputs is
labeled equivalent; sampling additional input sets is straightforward and left to
future work.
 
Once the divergence label is fixed, we generate a natural-language explanation of
the pair with \path{gemini-2.5-flash} through OpenRouter%
~\cite{openrouter2025} at temperature $0.2$. The prompt is grounded in a
structural diff of the two programs, which lists the concrete edits such as
renamed values, changed constants, substituted operations, and altered indexing,
so the model interprets known differences instead of inventing SSA identifiers.
These explanations ship with the dataset and are passed to the Judge Agent as
context, but they never determine the label: scoring always uses the
execution-derived divergence flag.
 
We use \path{iree-benchmark-module} to benchmark each program using a ten-second
warmup and thirty measured repetitions, and record mean latency, standard
deviation, and the p50, p90, and p99 percentiles. Linalg programs are benchmarked
on their embedded constants and torch programs on type-only inputs that IREE
zero-fills, under the same floating-point-deterministic flags used for labeling.
The complete pipeline is given in Algorithm 1 
(Appendix C).

\subsection{LLM Reasoning Agent}
\label{sec:reasoning_agent}

The LLM Reasoning Agent is the execution-hidden semantic evaluation harness of MLIRBench. It measures whether a model can determine the semantic relationship between two MLIR programs from the intermediate representation alone, with no access to compilation or runtime results. The evaluation protocol is controlled and fully logged,
allowing stored inputs and outputs to be audited.

For each program pair, the evaluated model observes only the original program $A$, the mutated program $B$, the dialect tag (\texttt{linalg} or \texttt{torch}), and the concrete deterministic inputs when they are available. All execution-derived information is withheld: the executed outputs of either program, the ground-truth semantic label, lowered IR, backend IR, execution traces, the compiler and runtime records produced during sandbox execution, and the Judge Agent scores. Under this isolation, a prediction must follow from reasoning over the source IR-SSA dataflow, tensor shapes and element types, affine indexing maps, iterator semantics, and dialect-specific operation behavior-rather than from any observation of how the programs behave when run.

The harness uses a predict-then-compare protocol. The model is instructed to predict the output behavior of programs $A$ and $B$ before classifying the pair with a single label $\hat{y} \in \{\textsc{Equiv}, \textsc{Diverge}, \textsc{Error}\}$. Separating per-program prediction from the pairwise decision keeps output-prediction errors distinguishable from classification errors. When concrete inputs are supplied, the model reasons over the same values used during sandbox execution, while the executed outputs remain hidden. The label space is restricted to these three classes, and the prompt does not permit abstention. A model that is unsure is instructed to commit to the best-supported label and lower its reported confidence accordingly.

Each response must be a single JSON object containing the predicted output for program $A$, the predicted output for program $B$, the label, a confidence score, the key semantic differences identified between the programs, an optional witness input or condition demonstrating divergence, and a natural-language explanation of the verdict. Responses are validated against this fixed schema before storage. Because models can emit malformed structured output, the parser performs lightweight recovery for formatting issues: it strips code fences, extracts the outer JSON object, sanitizes stray control characters, and validates the required fields and label value. For baseline runs, parsing or schema failures trigger one retry with a forced-verdict instruction that restates the required fields and demands a committed label. Unresolved failures retain their status for auditing. Gemini 3.1 Pro was evaluated without application-level retries. Under the scoring policy, unsuccessful predictions receive zero hard-label credit and are excluded from diagnostic score averages. Valid predictions are passed to the Judge Agent together with the concealed execution record for scoring.

Additional details on the prompt layout, visible and hidden artifacts, response schema, retry semantics, model roster, stored records, and reproducibility procedure are provided in Appendix D.

\subsection{Judge Agent}
\label{sec:judge_agent}

The Judge Agent is the final evaluation component of MLIRBench. Its role is to score the structured prediction produced by the LLM Reasoning Agent against the execution-grounded record produced by the Execution Sandbox Agent.

Each judge invocation receives the original MLIR program, the mutated MLIR program, dialect-specific guidance, mutation metadata, sandbox execution evidence, and the evaluated model's structured response. The sandbox record is treated as the sole source of semantic ground truth. If either program fails compilation, verification, or execution, the normalized label is \textsc{ERROR}. Otherwise, strict output comparison determines whether the pair is labeled \textsc{EQUIV} or \textsc{DIVERGE}. The evaluated model's predicted label is normalized into the same three-class space before scoring.

The Judge Agent returns three complementary scores. First, hard-label accuracy is computed deterministically as $\mathbb{1}[\hat{y}=y]$, where $\hat{y}$ is the model prediction and $y$ is the execution-derived label. This component does not depend on the judge model. Second, result-behavior accuracy $r \in [0,10]$ measures whether the model's predicted outputs or failure behavior are compatible with execution. For successful executions, the LLM applies $r=r_{\mathrm{structure}}+r_{\mathrm{exactness}}+3c/N$: $N$ counts reference scalar outputs across both programs, including omitted values, and $c$ counts correctly predicted values. Mutation metadata selects $p\in\{1,2,3,4\}$; structure earns $0$, $p/2$, or $p$, while exactness earns $7-p$ only when literals or unambiguous symbolic claims determine every output. For reference \textsc{ERROR} cases, a deterministic post-check assigns $r=10$ for recognized failure predictions and $r=0$ otherwise. Third, explanation quality $q \in [0,4]$ evaluates whether the explanation is causal, evidence-specific, dialect-aware, and consistent with the MLIR and sandbox evidence.

The aggregate score is computed as:
\[
S = \frac{100}{3}
\left(
\mathbb{1}[\hat{y}=y] + \frac{r}{10} + \frac{q}{4}
\right).
\]
Equal weighting is a reporting convention, not a calibrated measure of semantic understanding. Missing or invalid candidate responses receive zero label credit and no diagnostic scores; invalid judge outputs remain unscored for retry.

We use a single fixed judge model, \texttt{gpt-5.4-mini}, for all evaluated systems. Its prompt includes execution evidence and, when available, a stored generated reference narrative. On the same 100-pair sample, three fresh runs each remain within 1.15 aggregate points of the original saved model-level scores. The alternative judge assigns higher scores but preserves the top-three ordering across conditions; Appendix E.14 reports the comparison and limitations.

\section{Dataset Construction}
\label{sec:dataset_construction}

\begin{figure*}[t]
    \centering
    \includegraphics[width=\linewidth]{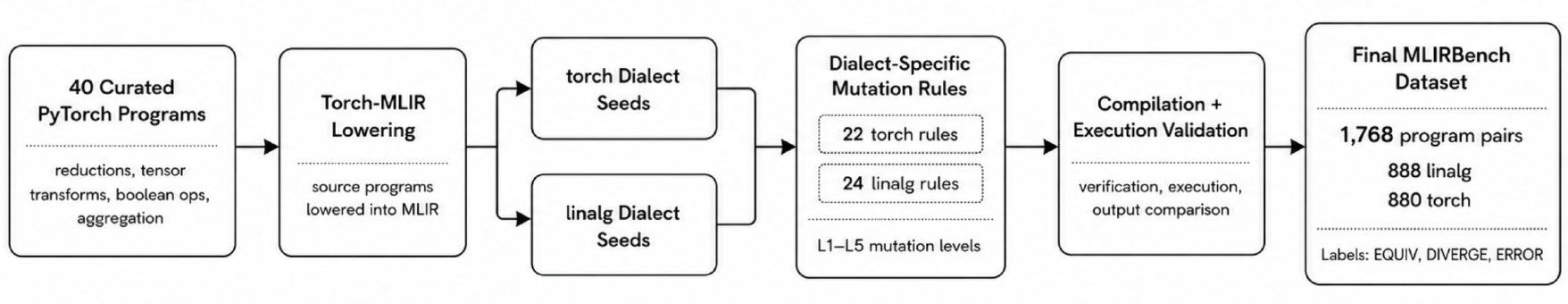}
    \caption{Dataset construction pipeline. MLIRBench starts from curated PyTorch programs, lowers them into \texttt{torch} and \texttt{linalg} MLIR, applies dialect-specific mutation rules, and validates the resulting program pairs through compilation and execution.}
    \label{fig:dataset_construction}
\end{figure*}

\subsection{Source Program Collection}

MLIRBench begins with \textbf{40 manually curated PyTorch programs} covering reductions, tensor transformations, aggregation, boolean operations, and shape-related computations. These programs are intentionally compact so that the semantic effect of each mutation can be isolated and analyzed. The final benchmark expands these seeds into \textbf{1,768 execution-validated MLIR program pairs}, with \textbf{888} pairs in \texttt{linalg} and \textbf{880} pairs in \texttt{torch}. Additional source-program and seed-category details are provided in Appendix A.1.

\subsection{Lowering PyTorch to MLIR}

Each PyTorch program is lowered using the Torch-MLIR compilation pipeline into two MLIR dialects: \texttt{torch} and \texttt{linalg}. The \texttt{torch} dialect preserves a representation closer to the original PyTorch program, while \texttt{linalg} exposes structured tensor computation, indexing behavior, and explicit dataflow.

After lowering, MLIRBench applies dialect-specific mutation rules across five mutation levels. In total, the dataset uses \textbf{46 mutation rules}: \textbf{24} for \texttt{linalg} and \textbf{22} for \texttt{torch}. The rules include SSA renaming, constant changes, operand substitution, operator replacement, structural rewrites, and failure-inducing mutations. Each generated pair is validated through verification, compilation, execution, and output comparison, then labeled as \textsc{EQUIV}, \textsc{DIVERGE}, or \textsc{ERROR} based on observed behavior.

Further details on the Torch-MLIR lowering process are provided in Appendix A.2, while mutation-rule coverage is summarized in Appendix B.3.

\begin{table}[t]
    \centering
    \caption{Summary of the MLIRBench dataset.}
    \label{tab:dataset_summary}
    \begin{tabular}{lr}
        \toprule
        Quantity & Count \\
        \midrule
        Source PyTorch programs & 40 \\
        Total MLIR program pairs & 1,768 \\
        \texttt{linalg} program pairs & 888 \\
        \texttt{torch} program pairs & 880 \\
        \texttt{linalg} mutation rules & 24 \\
        \texttt{torch} mutation rules & 22 \\
        Total mutation rules & 46 \\
        \bottomrule
    \end{tabular}
\end{table}

\section{Results}

\subsection{Model Reasoning Results}
\label{sec:model_reasoning_results}

\label{sec:benchmark_results}
\begin{figure*}[t]
    \centering
    \includegraphics[width=\textwidth]{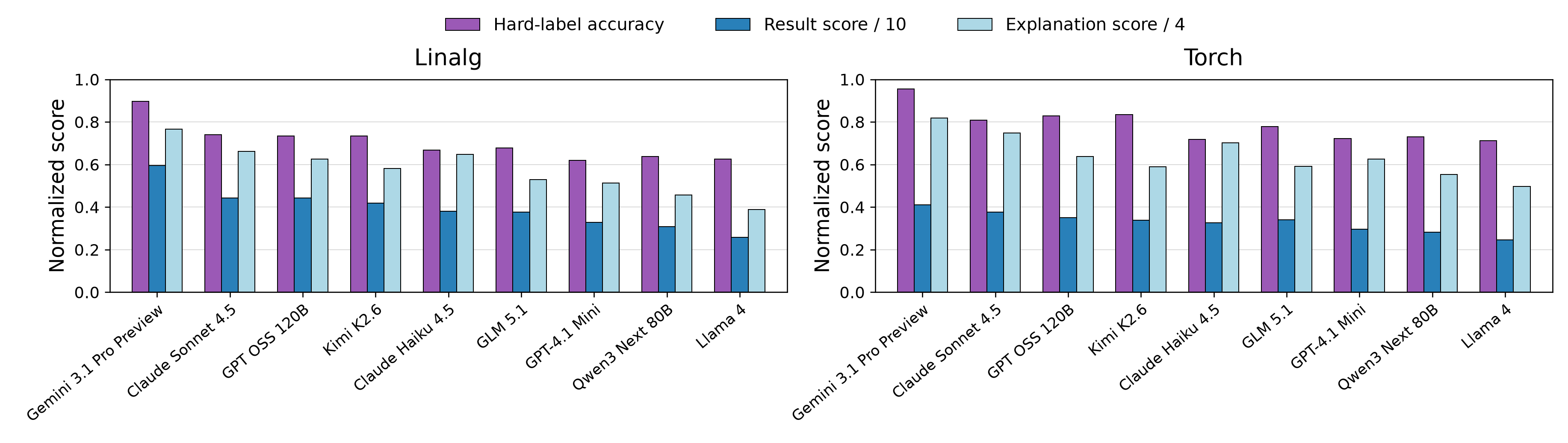}
    \caption{Benchmark performance across nine evaluated models on 888 \texttt{linalg} and 880 \texttt{torch} pairs. Label accuracy includes all attempts; result and explanation scores are normalized by their maxima and averaged over valid responses.}
    \label{fig:9-model-benchmark}
\end{figure*}

The LLM Reasoning Agent was evaluated over the MLIR program-pair dataset produced by the Execution Sandbox Agent, with all execution artifacts withheld from the evaluated models. The final evaluation covers 888 \texttt{linalg} pairs and 880 \texttt{torch} pairs, for a total of 1{,}768 program pairs. Each pair is evaluated independently by nine models, producing 15{,}912 structured reasoning records. Each record stores the model identifier, dialect tag, pair metadata, predicted outputs for programs $A$ and $B$, predicted semantic label, confidence score, explanation, retry count, latency, and parsing status. Appendix D expands the prompt structure, response schema, retry logic, model roster, and stored record format.

The protocol is controlled and fully logged. Every model receives
the same prompt template, dialect information, and deterministic
input values for each program pair. All models are queried at
temperature 0, and provider model identifiers are recorded.
The retained structured records support auditing without rerunning
the benchmark. Table~\ref{tab:reasoning_scale} reports the
evaluation scale.

\begin{table}[t]
\centering
\caption{Reasoning Agent evaluation scale.}
\label{tab:reasoning_scale}
\begin{tabular}{lr}
\toprule
Quantity & Value \\
\midrule
\texttt{linalg} program pairs & 888 \\
\texttt{torch} program pairs  & 880 \\
Total program pairs           & 1{,}768 \\
Models evaluated              & 9 \\
Structured reasoning records  & 15{,}912 \\
\bottomrule
\end{tabular}
\end{table}

Each response is parsed against a fixed JSON schema before storage. Responses that fail parsing or schema validation are retried once with a forced-verdict instruction that restates the required fields and requires a committed label. After validation, the retained structured responses are passed to the scoring stage together with the hidden execution record, which supplies the ground-truth label and serves as the sole reference for evaluating each prediction.

\begin{table}[t]
\centering
\caption{Reasoning Agent scoring summary. Hard-label accuracy is computed deterministically against the execution-derived label; result-behavior accuracy measures the predicted concrete output behavior, and explanation quality measures whether the explanation is causal, dialect-aware, and grounded in the MLIR evidence.}
\label{tab:reasoning_scoring}
\begin{tabular}{lr}
\toprule
Quantity & Value \\
\midrule
Structured reasoning records  & 15{,}912 \\
Mean aggregate score          & 57.21 \\
Mean hard-label accuracy      & 74.59\% \\
Mean result-behavior accuracy & 3.63 / 10 \\
Mean explanation-quality score & 2.43 / 4 \\
\bottomrule
\end{tabular}
\end{table}

The results expose a gap between two distinct abilities. Hard-label accuracy, which measures whether a model assigns the correct \textsc{Equiv}, \textsc{Diverge}, or \textsc{Error} label, is substantially higher than result-behavior accuracy, which measures whether the model correctly predicts what the two programs compute. A model can therefore often classify a pair correctly without accurately tracing the concrete behavior that justifies the classification. This distinction matters for compiler IR reasoning because a correct label alone does not establish that a model has followed the underlying dataflow, shapes, tensor values, or failure mode. Because the label prediction and the behavioral prediction are scored separately against the same execution record, MLIRBench makes this gap measurable rather than collapsing it into a single accuracy number.

\subsection{Benchmark}

Figure ~\ref{fig:9-model-benchmark} summarizes model performance. The main trend is a clear gap between hard-label accuracy and execution-level reasoning. Models often predict the correct \textsc{EQUIV}, \textsc{DIVERGE}, or \textsc{ERROR} label, but struggle to predict the concrete behavior that supports that label.

Gemini 3.1 Pro~\cite{google2026gemini31pro} achieves the highest aggregate score at 74.18, followed by Claude Sonnet 4.5~\cite{anthropic2025sonnet45}  and GPT-OSS 120B~\cite{openai2025gptoss}. Gemini obtains a judge-assigned result-behavior score of 5.04/10 and an explanation-quality score of 3.17/4. Its hard-label accuracy is 92.65\% across all 1,768 benchmark pairs when the five unsuccessful predictions are counted as incorrect. Among scored predictions, Gemini correctly classifies 97.02\% of \textsc{ERROR} cases but only 68.73\% of \textsc{DIVERGE} cases, including 84 divergent pairs incorrectly classified as equivalent. These results suggest that detailed output prediction remains challenging despite strong semantic classification performance.

Table~\ref{tab:level_results} combines the \texttt{linalg} and
\texttt{torch} mutation families into five mutation levels. Performance
varies across mutation levels and models. Most models show substantial
difficulty on Level~5 failure-inducing mutations, while Gemini 3.1 Pro
performs strongly on these cases. Its main remaining weakness is behavioral
divergence, correctly classifying 68.73\% of \textsc{DIVERGE} pairs.

\begin{table}[h]
  \caption{Average benchmark score by mutation level, combining \texttt{linalg} and \texttt{torch}.}
  \label{tab:level_results}
  \begin{tabular}{lrrrrr}
    \toprule
    Model & L1 & L2 & L3 & L4 & L5 \\
    \midrule
    Gemini 3.1 Pro & 68.16 & 68.58 & 72.82 & 67.57 & 94.94 \\
    Claude Sonnet 4.5 & 63.82 & 60.72 & 65.38 & 65.13 & 59.11 \\
    GPT-OSS 120B & 60.90 & 61.08 & 65.02 & 62.47 & 52.59 \\
    Kimi K2.6 & 57.48 & 56.52 & 63.62 & 62.11 & 51.67 \\
    Claude Haiku 4.5 & 59.14 & 56.70 & 56.29 & 60.54 & 53.57 \\
    GLM 5.1 & 59.34 & 58.40 & 61.50 & 61.69 & 32.78 \\
    GPT-4.1 Mini & 61.23 & 55.90 & 53.46 & 59.82 & 26.88 \\
    Qwen3 Coder Next & 56.62 & 52.21 & 58.23 & 57.95 & 21.18 \\
    Llama 4 & 52.42 & 49.36 & 52.54 & 53.63 & 18.15 \\
    \bottomrule
  \end{tabular}
\end{table}

Overall, MLIRBench distinguishes strong semantic classification
from detailed output prediction. Gemini demonstrates strong
compiler-error recognition, but still misses behavior-changing
mutations and receives limited result-behavior credit. These
remaining gaps are important for compiler transformation validation
and debugging.

\subsection{Dataset}

MLIRBench releases the original and mutated MLIR programs together with
dialect and mutation metadata, compiler and execution status, execution-derived
semantic labels, runtime measurements, model predictions and explanations,
and Judge Agent outputs. These artifacts support both benchmark auditing and
downstream compiler-reasoning experiments. Dataset construction details are
provided in Appendix A, mutation records in Appendix B, execution records in
Appendix C, and additional analysis in Appendix F. \\

\noindent \textbf{Dataset Availability:} The MLIRBench dataset, including original and mutated MLIR programs, execution-derived labels, mutation metadata, and evaluation artifacts, is publicly available at \url{https://huggingface.co/datasets/mlirbench/MLIRBench}.

\begin{table}[t]
\centering
\caption{Artifacts provided by MLIRBench.}
\label{tab:dataset_contents}
\begin{tabular}{ll}
\toprule
\textbf{Category} & \textbf{Included} \\
\midrule
Programs & Original and mutated MLIR pairs \\
Dialects & \texttt{torch}, \texttt{linalg} \\
Annotations & Mutation level and metadata \\
Execution Data & Compile, runtime, and labels \\
Performance Data & Latency and percentile statistics \\
Model Outputs & Predictions and explanations \\
Evaluation Data & Judge scores and benchmark metrics \\
\bottomrule
\end{tabular}
\end{table}

\section{Discussion}

\subsection{Mutation Agent}

A major challenge in constructing MLIRBench was generating mutations that were both compiler-valid and faithful to the intended mutation rule. Unlike ordinary code generation, each mutation must preserve the intended scope of change, satisfy dialect-specific constraints, and avoid unrelated semantic edits that would make the resulting program pair ambiguous. During development, we tested several open-source models for mutation generation, including \texttt{gpt-oss:20b}, \texttt{gpt-oss:120b}, \texttt{qwen2.5} variants, \texttt{qwen3-coder:480b}, and \texttt{qwen3-coder-next}. These models often produced syntactically plausible MLIR, but they were inconsistent under strict mutation constraints. Common failures included hallucinated operations, invalid SSA references, malformed affine maps, tensor shape mismatches, and edits that extended beyond the requested mutation level.

These issues became more pronounced for structural and failure-inducing mutations, where small errors could create invalid benchmark instances or compiler failures unrelated to the target transformation. Regenerating multiple candidates improved quality in some cases, but was not reliable enough for large-scale benchmark construction. We therefore used GPT-5.5 for mutation generation, which followed mutation requirements more consistently and better preserved constraints in both the \texttt{torch} and \texttt{linalg} dialects. Even with this stronger generator, validation remained necessary: every mutation is checked before entering MLIRBench. This shows that controlled compiler transformation generation is harder than producing syntactically valid IR, and that validation-guided construction is essential for reliable compiler-reasoning datasets.

\subsection{Execution Sandbox Agent}
\label{sec:execution_experiments}

The released benchmark contains 1{,}768 program pairs: 888 \texttt{linalg}
and 880 \texttt{torch}. One L2 linalg parameter-mutation rule was
excluded after producing invalid MLIR across all 37 programs. Of the retained
pairs, 1{,}427 executed successfully; 1{,}402 also have complete benchmark
timing, while 25 torch T2 pairs were retained as runtime errors due to altered
input signatures. The remaining records include 338 L5/T5
\texttt{expected\_invalid} pairs and 3 linalg compilation failures.
Table~\ref{tab:divergence} summarizes the 1{,}765 labeled pairs.

\begin{table}[t]
\centering
\caption{Behavioral divergence by mutation level over 1{,}765 labeled pairs.
\textsc{Diverge} counts output mismatches under bit-exact equality; ExpInv
denotes expected-invalid L5/T5 records.}
\label{tab:divergence}
\small
\begin{tabular}{llrrrr}
\toprule
Dialect & Level & Total & Diverge & Equiv. & ExpInv \\
\midrule
Linalg & L1 & 185 & 17 (9\%)  & 168 (91\%) & 0   \\
Linalg & L2 & 147 & 76 (52\%) & 71 (48\%)  & 0   \\
Linalg & L3 & 184 & 41 (22\%) & 143 (78\%) & 0   \\
Linalg & L4 & 184 & 32 (17\%) & 152 (83\%) & 0   \\
Linalg & L5 & 185 & 0 (0\%)   & 7 (4\%)    & 178 \\
\midrule
Torch  & T1 & 200 & 3 (2\%)   & 197 (98\%) & 0   \\
Torch  & T2 & 160 & 88 (55\%)$^\dagger$ & 72 (45\%) & 0 \\
Torch  & T3 & 160 & 10 (6\%)  & 150 (94\%) & 0   \\
Torch  & T4 & 200 & 8 (4\%)   & 192 (96\%) & 0   \\
Torch  & T5 & 160 & 0 (0\%)   & 0 (0\%)    & 160 \\
\midrule
\multicolumn{2}{l}{\textbf{Total}} & \textbf{1{,}765}
  & \textbf{275 (16\%)} & \textbf{1{,}152 (65\%)} & \textbf{338 (19\%)} \\
\bottomrule
\end{tabular}
\\[3pt]
{\footnotesize $^\dagger$ 25 T2 divergences involve altered function input
signatures; among the remaining 135 matched-input pairs, 47\% diverge.}
\end{table}

Level-2 mutations show the highest divergence rates at 52\% for
\texttt{linalg} and 55\% for \texttt{torch}, while structural mutations
range from 4\% to 22\%. Failure-inducing levels behave as intended, with
96\% of L5 and 100\% of T5 pairs failing compilation or validation.
Across 1{,}402 timed pairs, the mean runtime delta is $-0.001$ ms with
standard deviation $0.020$ ms.

As an execution-contract baseline, MLIRBench uses bit-exact output equality.
We compare this baseline against two tolerance-based variants by recomputing
the execution labels from the stored outputs using \texttt{numpy.allclose}.
We test NumPy's default tolerances,
$(r_{\mathrm{tol}},a_{\mathrm{tol}})=(10^{-5},10^{-8})$, and a looser
sensitivity setting of $(10^{-4},10^{-6})$. Neither setting changes any of
the 275 \textsc{Diverge} labels among the 1{,}427 successfully executed
benchmark pairs.

\subsection{LLM Reasoning Agent}

The results show that hard-label accuracy alone does not establish
detailed semantic understanding. A model may correctly predict
\textsc{Equiv}, \textsc{Diverge}, or \textsc{Error} without predicting
the concrete behavior that justifies the label. MLIRBench keeps
these abilities separate by scoring hard-label accuracy independently
from result-behavior accuracy and explanation quality.

This distinction matters because exact MLIR reasoning requires
tracking SSA dependencies, tensor shapes, element types, affine
indexing maps, iterator semantics, constants, return values, and
verifier constraints at the same time. A local edit can propagate
through the dataflow graph in ways that are not obvious from surface
syntax. Failure-inducing mutations require compiler-legality reasoning
over type compatibility, shape relations, operand constraints, and
dependency well-formedness, but their difficulty varies across models.
Thus, models that recognize semantic relationships or compiler errors
may still struggle with detailed output prediction, which is important
for transformation validation, failure diagnosis, and optimization
assistance.

\subsection{Judge Agent}

The Judge Agent is useful because label correctness alone does not capture whether a model understands the compiler-level cause of a result. A model may correctly predict \textsc{Error} while missing the invalid operand type, broken SSA dependency, illegal shape relation, or dialect-specific verifier rule. Conversely, it may produce a fluent explanation that cites operations or constraints not present in the program. By grounding scoring in both the MLIR pair and the execution record, MLIRBench penalizes unsupported reasoning rather than only incorrect labels.

The main limitation is scorer variance in LLM-based explanation grading. We reduce this risk by computing hard-label accuracy deterministically, using a fixed judge model for all systems, requiring structured JSON outputs, and applying deterministic post-checks for error recognition. Explanation-quality scores should therefore be interpreted as diagnostic signals rather than absolute measures of semantic understanding.

\raggedbottom
\subsection{Limitations}

\begin{enumerate}
\item \textbf{Limited Dialect Coverage:}
MLIRBench currently focuses on the \texttt{torch} and \texttt{linalg} dialects. While these dialects capture both framework-level and structured tensor representations, the benchmark does not yet cover other widely used MLIR dialects such as \texttt{affine}, \texttt{scf}, \texttt{vector}, \texttt{gpu}, \texttt{llvm}, or \texttt{memref}.

\item \textbf{Limited Source-Program Diversity:}
The benchmark originates from 40 manually curated PyTorch programs. Although mutation generation substantially expands the dataset, the underlying workload diversity remains bounded by the selected seed programs.

\item \textbf{Execution Environment Dependence:}
By using IREE ground-truth labels are generated under a fixed compilation and execution configuration. Results may vary across alternative MLIR backends, or hardware targets.

\item \textbf{Controlled Rather Than Real-World Transformations:}
MLIRBench is constructed from systematically generated mutations designed to isolate specific semantic changes. Real-world compiler optimizations and developer-authored transformations may exhibit additional complexity beyond the current benchmark.

\item \textbf{Structured-Output Reliability:}
Two \texttt{gpt-oss-120b} responses failed the required JSON response contract during evaluation: one produced a \texttt{PARSE\_ERROR} and another a \texttt{SCHEMA\_ERROR}. These were model-interface formatting failures rather than issues with the MLIR pairs or execution-derived labels, and suggest using stricter schema-constrained generation in future versions of the harness.
 
\end{enumerate}

\section{Downstream Tasks}

\subsection{Task 1: Compiler-Grounded Optimization-Benefit Prediction}

The first downstream task evaluates whether MLIRBench can support
compiler-aware optimization reasoning. Given an MLIR program, the model
predicts the benefit of a real compiler transformation using only the original
program and its static structure. We define one task for each dialect:
elementwise producer--consumer fusion for \texttt{linalg} using
\path{linalg-fuse-elementwise-ops}, and the
\path{torch-simplification-pipeline} for \texttt{torch}. Compiler outputs are
used to assign \textsc{Low}, \textsc{Medium}, or \textsc{High} benefit labels,
while the transformed IR and measured result are hidden from the model.

After compiler validation, the task contains 693 Linalg and 687 Torch
programs. We use family-disjoint five-fold evaluation and compare zero-shot,
three-shot, and LoRA-based SFT. Since the classes are imbalanced, macro-F1 is
used as the main metric.

Figure~\ref{fig:dt1_supervision} shows the results. For Linalg fusion,
macro-F1 increases from 26.8\% zero-shot and 24.1\% three-shot to
\textbf{60.9\%} after SFT. For Torch simplification, it increases from
9.3\% and 22.1\% to \textbf{48.0\%}. The compiler transformations modify
91.8\% of compiler-valid Linalg programs and 45.0\% of Torch programs, with
the labels derived directly from observed compiler transformations.

{
\setlength{\intextsep}{3pt}

\begin{figure}[!htbp]
    \centering
    \includegraphics[width=\linewidth]
    {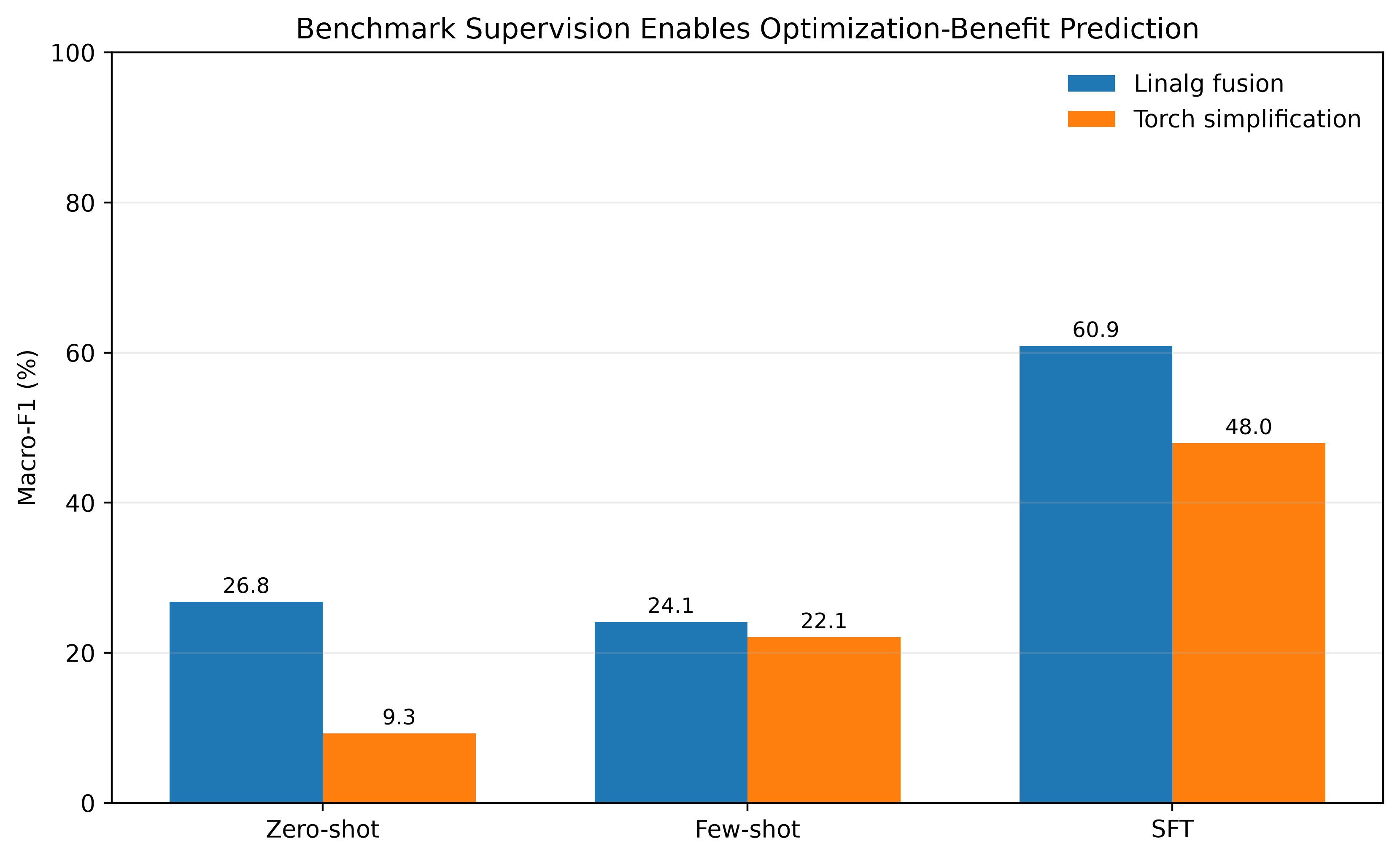}
    \caption{Optimization-benefit prediction for Linalg fusion and Torch
    simplification. Fine-tuning on MLIRBench-derived supervision improves
    macro-F1 over zero-shot and few-shot prompting for both dialects.}
    \label{fig:dt1_supervision}
\end{figure}
}

\begin{figure}[t]
    \centering
    \includegraphics[width=0.72\linewidth]
    {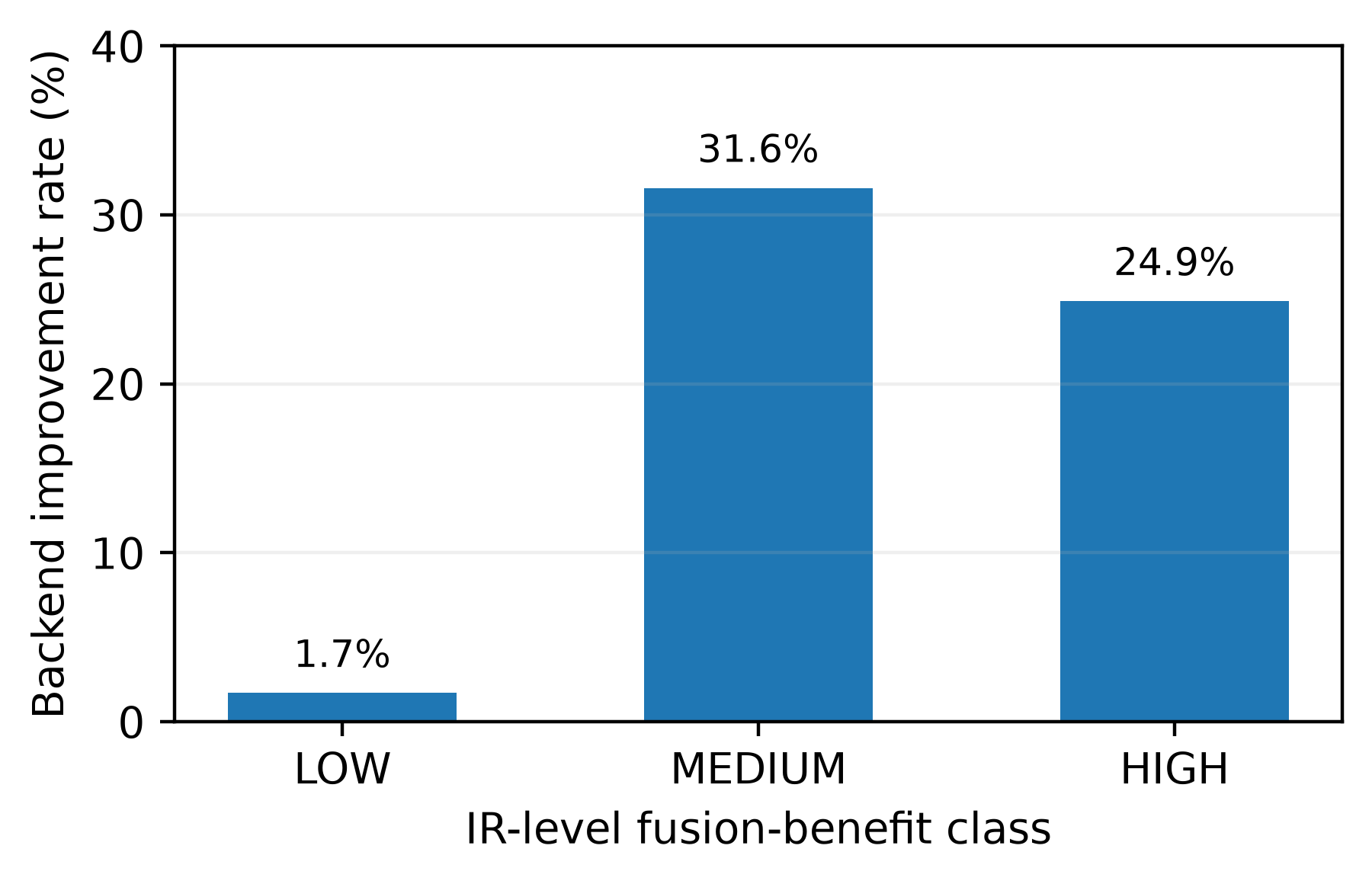}
    \caption{RISC-V backend improvement rate after Linalg fusion, grouped by
    IR-level fusion-benefit class.}
    \label{fig:dt1_riscv}
\end{figure}

We also evaluate how Linalg fusion relates to RISC-V code generation by
compiling the original and fusion-transformed programs and comparing their
static instruction streams. Backend improvement is observed in 1.7\% of
\textsc{Low}, 31.6\% of \textsc{Medium}, and 24.9\% of \textsc{High} cases.
As shown in Figure~\ref{fig:dt1_riscv}, medium- and high-benefit programs are
substantially more likely to show backend improvement than low-benefit
programs. Overall, Task~1 shows that MLIRBench supports compiler-grounded
optimization-benefit prediction across both dialects, with additional
RISC-V-aware analysis.

\subsection{Task 2: Transfer of Execution-Grounded Semantic Supervision}

The second downstream task studies whether MLIRBench's semantic labels can
provide supervision for unseen program families. Each example contains an
original MLIR program and a mutated version, with one of three
execution-grounded labels: \textbf{EQUIV}, \textbf{DIVERGE}, or
\textbf{ERROR}, representing preserved behavior, changed behavior, or
verification/compilation/execution failure, respectively.

We use \path{Qwen2.5-Coder-14B-Instruct} model with family-disjoint train, validation, and test splits. We compare zero-shot and balanced six-shot prompting with LoRA-based SFT to study whether MLIRBench supervision generalizes beyond the examples seen during training or prompting.

\begin{figure}[t]
    \centering
    \includegraphics[width=\linewidth]
    {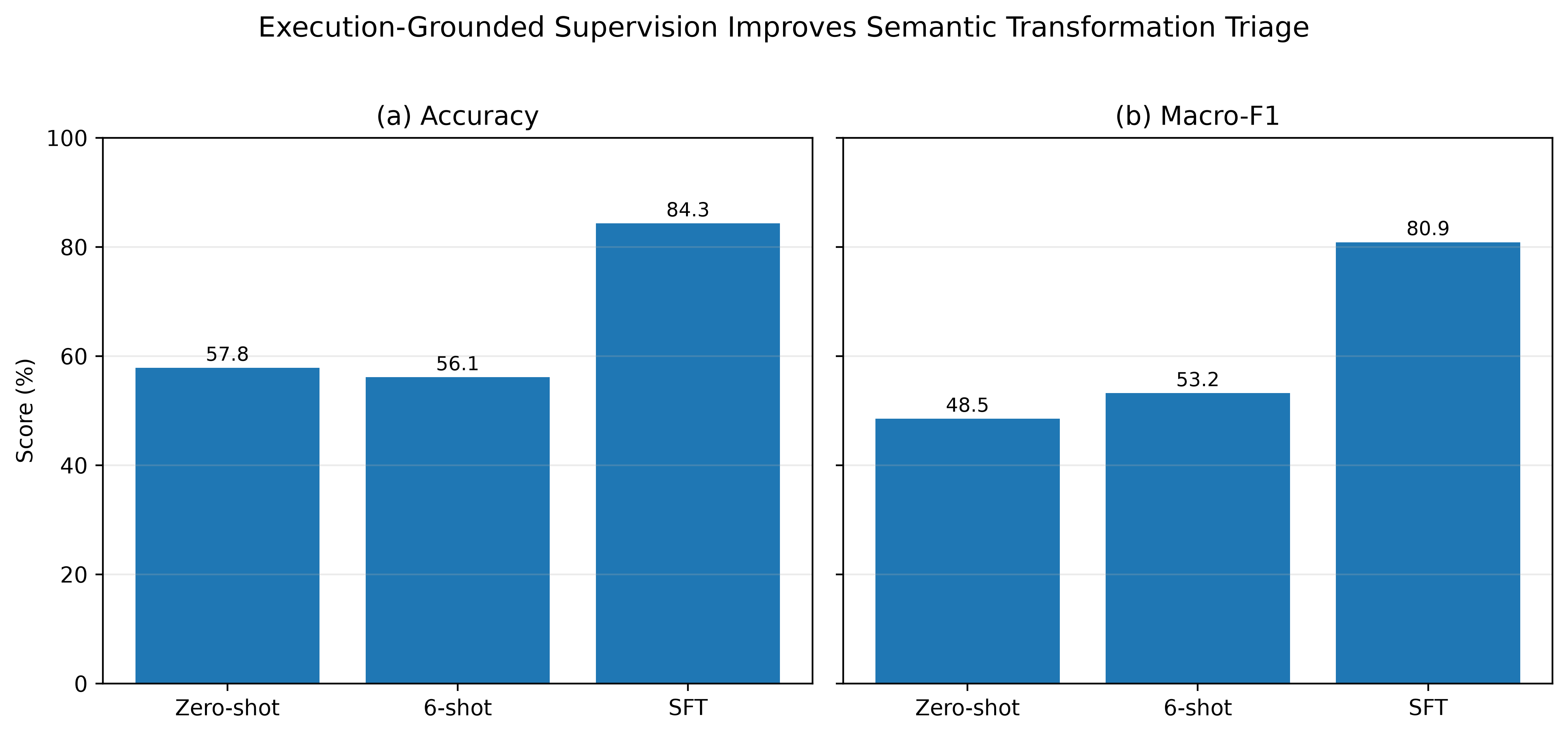}
    \caption{Semantic classification on family-disjoint MLIRBench program
    pairs. MLIRBench supervision improves both accuracy and macro-F1 on
    held-out program families.}
    \label{fig:dt2_supervision}
\end{figure}

Figure~\ref{fig:dt2_supervision} summarizes the results. Zero-shot prompting
achieves 57.8\% accuracy and 48.5\% macro-F1, while six-shot prompting
achieves 56.1\% and 53.2\%. SFT reaches \textbf{84.3\% accuracy} and
\textbf{80.9\% macro-F1} on held-out program families. The improvement is
consistent across both dialects, reaching 80.8\% macro-F1 on \texttt{linalg}
and 80.4\% on \texttt{torch}.

We additionally examine false-equivalence predictions, where a
\textit{DIVERGE} or \textit{ERROR} pair is incorrectly classified as
\textit{EQUIV}. This rate decreases from 59.1\% zero-shot to 27.4\%
six-shot and \textbf{24.1\%} after SFT. The final model achieves per-class
F1 scores of 88.1\% for \textit{EQUIV}, 59.7\% for \textit{DIVERGE}, and
94.8\% for \textit{ERROR}.

Overall, these results show that MLIRBench captures the execution-grounded semantic information that transfers to unseen program families across both
\texttt{linalg} and \texttt{torch}. Full prompt, training, and evaluation
details are provided in Appendix G.

\section{Conclusion}

We introduced MLIRBench, an execution-grounded benchmark and dataset for
evaluating semantic reasoning over MLIR programs. MLIRBench combines
dialect-aware mutation generation, deterministic compilation and execution,
execution-hidden model prediction, and structured scoring across
\texttt{torch} and \texttt{linalg}. The resulting dataset contains 1,768
execution-validated program pairs spanning five mutation levels and three
semantic outcomes: \textsc{EQUIV}, \textsc{DIVERGE}, and \textsc{ERROR}.

Our evaluation shows that current LLMs can often identify the high-level
semantic relationship between MLIR programs, but exact execution reasoning
remains considerably harder. Gemini 3.1 Pro achieves the strongest overall
performance with an aggregate score of 74.18 and 92.65\% hard-label accuracy,
while its result-behavior score reaches 5.04/10. Performance also varies across
mutation types and models, with behavioral divergence, structural changes, and
compiler-legality constraints exposing important reasoning failures. These
results show why execution-grounded evaluation is useful: label accuracy alone
does not capture whether a model correctly understands the behavior that
produces that label.

The downstream experiments further show that MLIRBench can provide useful
supervision beyond benchmark evaluation. Execution-grounded semantic
supervision reaches 84.3\% accuracy and 80.9\% macro-F1 on held-out program
families, with similar performance across \texttt{linalg} and \texttt{torch}.
For compiler optimization reasoning, MLIRBench-derived supervision improves
optimization-benefit prediction for both Linalg fusion and Torch
simplification, while the RISC-V backend analysis connects the predicted
IR-level benefit to downstream code generation. Together, these results show
that MLIRBench supports both semantic reasoning and compiler-grounded
optimization analysis across multiple MLIR representations.

The benchmark remains limited to two dialects and 40 source programs, and its
execution labels are produced under a fixed compiler configuration. Future
work will expand dialect and workload coverage, evaluate additional compiler
backends and hardware targets, and use the benchmark's structured execution
and failure signals to develop more capable compiler-reasoning models.

\begin{acks}
This work was supported by the Gemini Academic Program Award, which provided access to Gemini models and computational resources that supported the experiments conducted in this research. The research was also partially supported by the HPC@ISU equipment at Iowa State University, some of which has been purchased through funding provided by NSF under MRI grant numbers 1726447 and 2018594. We also thank Dr. Simanta Mitra for his guidance and support during our Senior Design course at Iowa State University and for facilitating our involvement in this project.
\end{acks}

\nocite{*}
\bibliographystyle{ACM-Reference-Format}
\bibliography{references}

\newpage

\appendix
\section*{Appendix}

\renewcommand\footnotetextcopyrightpermission[1]{}
\settopmatter{printacmref=false}


\raggedbottom

\setlength{\textfloatsep}{6pt plus 1pt minus 1pt}
\setlength{\floatsep}{5pt plus 1pt minus 1pt}
\setlength{\intextsep}{5pt plus 1pt minus 1pt}
\setlength{\dbltextfloatsep}{6pt plus 1pt minus 1pt}
\setlength{\dblfloatsep}{5pt plus 1pt minus 1pt}

\appendix

\section{Dataset Construction}

This appendix provides additional details regarding the construction of the MLIRBench dataset, including source program selection and the process used to generate MLIR representations.

\subsection{Source Programs}

The foundation of MLIRBench is a collection of manually curated PyTorch programs designed to capture common tensor computation patterns while remaining sufficiently compact for systematic mutation and execution analysis. Rather than relying on large application workloads, we intentionally selected small, self-contained programs inspired by educational programming exercises and numerical computing tasks. This design allows semantic changes introduced by mutations to be isolated and analyzed without interference from unrelated program complexity.

We collected a total of 40 PyTorch programs and organized them into four seeds, with 10 programs per seed. Each seed was designed to represent a distinct category of tensor computation commonly encountered in machine learning, scientific computing, and tensor-processing workloads.

\paragraph{Seed 1: Reduction Operations}
Programs in this category reduce multi-dimensional tensors to scalar outputs through aggregation operations. Representative examples include \texttt{sum\_2d\_tensor}, \texttt{sum\_3d\_tensor}, \texttt{max\_value}, \texttt{min\_value}, \\ \texttt{mean\_reduction}, \texttt{variance\_reduction}, \texttt{count\_nonzero}, and \texttt{count\_above\_threshold}.

\paragraph{Seed 2: Small Tensor Operations}
This category focuses on structured tensor transformations and arithmetic operations producing tensor-valued outputs. Representative programs include \texttt{matmul\_2x2}, \texttt{matmul\_3x3}, \\ \texttt{elementwise\_add\_small}, \texttt{elementwise\_mul\_small}, \\\texttt{transpose\_small}, and \texttt{reshape\_small}.

\paragraph{Seed 3: Boolean Reasoning Operations}
Programs in this seed generate boolean outputs and require reasoning about logical properties of tensors. Representative tasks include \texttt{all\_positive}, \texttt{any\_negative}, \texttt{tensors\_equal}, \texttt{tensors\_close}, \texttt{contains\_nan}, and \texttt{is\_sorted}.

\paragraph{Seed 4: Aggregation Operations}
This category captures workloads that reduce tensors along selected dimensions while preserving partial tensor structure. Examples include \texttt{sum\_along\_axis}, \texttt{max\_per\_row}, and \texttt{softmax\_small}.

Together, these four seeds provide coverage of arithmetic, structural, logical, and aggregation-oriented computation patterns while maintaining manageable program sizes suitable for large-scale mutation generation and execution-based validation.

\begin{table}[t]
\centering
\caption{Source program seed categories used in MLIRBench.}
\label{tab:app_seed_categories}
\small
\begin{tabular}{llr}
\toprule
Seed & Category & Programs \\
\midrule
Seed 1 & Reduction operations & 10 \\
Seed 2 & Small tensor operations & 10 \\
Seed 3 & Boolean reasoning operations & 10 \\
Seed 4 & Aggregation operations & 10 \\
\midrule
Total & -- & 40 \\
\bottomrule
\end{tabular}
\end{table}

\subsection{Lowering to MLIR}

One challenge during dataset construction was the lack of publicly available MLIR datasets suitable for large-scale semantic reasoning studies. Existing MLIR repositories primarily consist of compiler tests, transformation examples, and dialect-specific demonstrations rather than diverse collections of execution-ready programs.

To address this limitation, we generated our own MLIR corpus by lowering curated PyTorch programs using the Torch-MLIR compilation pipeline. Starting from PyTorch programs provided semantically meaningful workloads with known behavior while enabling systematic dataset construction from a widely used machine-learning framework.

Each benchmark program was implemented as a PyTorch module and imported into MLIR using the Torch-MLIR FX importer with representative inputs. This process produced programs in the \texttt{torch} dialect, preserving framework-level semantics while exposing computations in MLIR form.

The resulting programs were subsequently lowered to the \texttt{linalg} dialect. While the \texttt{torch} dialect retains higher-level operator semantics, the \texttt{linalg} dialect exposes structured tensor computations, affine indexing behavior, iterator semantics, and explicit dataflow relationships.

The final corpus contains paired \texttt{torch} and \texttt{linalg} representations for each source program. These dialect-specific programs serve as the seeds for mutation generation, benchmark construction, execution-grounded evaluation, and downstream learning tasks.

\section{Mutation Agent: Implementation Details}
\label{app:mutation-agent}

This appendix provides implementation details for the Mutation Agent used in MLIRBench. The main paper describes the role of the agent in the benchmark pipeline; here we document the mutation inputs, dialect-specific analysis, mutation levels, validation procedure, stored metadata, and common generation failures.

\subsection{Mutation Inputs}
\label{app:mutation-inputs}

The Mutation Agent receives an MLIR seed program, a dialect tag, a mutation level, and a mutation requirement. The dialect tag is either \texttt{torch} or \texttt{linalg}. The mutation level specifies the intended difficulty of the transformation, while the mutation requirement describes the specific edit to apply. The output is a candidate pair consisting of the original MLIR program and a mutated version.

Each mutation request is designed to produce one controlled program-pair instance. The goal is not to generate arbitrary MLIR, but to apply a targeted transformation while preserving the rest of the program as much as possible. This allows the benchmark to isolate the semantic effect of each mutation during execution-based validation.

\subsection{Dialect-Specific Program Analysis}
\label{app:mutation-analysis}

Before generating a mutation, the agent performs dialect-specific inspection of the input program. For \texttt{torch}, the analysis focuses on ATen operators, tensor shapes, scalar constants, dimension arguments, dtype annotations, broadcasting behavior, and SSA dependencies. For \texttt{linalg}, the analysis focuses on structured operations, indexing maps, iterator types, region bodies, yielded values, operand and result types, and destination-style semantics.

The extracted metadata is used to guide the mutation prompt. This reduces invalid edits by giving the generation model information about the structure of the program before it attempts to modify it. The metadata also helps constrain the mutation so that the generated program changes the intended part of the IR rather than introducing unrelated edits.

\begin{table}[t]
\centering
\caption{Mutation levels used by the Mutation Agent.}
\label{tab:app_mutation_levels}
\small
\begin{tabular}{p{0.12\columnwidth}p{0.76\columnwidth}}
\toprule
Level & Description \\
\midrule
L1 & Syntax-preserving edits such as SSA renaming and formatting changes \\
L2 & Parameter or attribute edits such as constants, dimensions, and dtypes \\
L3 & Operator substitutions between related operations \\
L4 & Structural edits affecting dataflow, shape behavior, or computation structure \\
L5 & Failure-inducing edits that violate compiler or runtime constraints \\
\bottomrule
\end{tabular}
\end{table}

\subsection{Mutation Levels and Rule Coverage}
\label{app:mutation-levels}

The mutation rules are dialect-specific because \texttt{torch} and \texttt{linalg} expose different operations and verifier constraints. In total, MLIRBench uses 46 mutation rules: 24 for \texttt{linalg} and 22 for \texttt{torch}. Table~\ref{tab:app_mutation_rules} summarizes the rule coverage.

\begin{table}[t]
\centering
\caption{Mutation rule coverage by dialect.}
\label{tab:app_mutation_rules}
\small
\begin{tabular}{lr}
\toprule
Dialect & Number of rules \\
\midrule
\texttt{linalg} & 24 \\
\texttt{torch} & 22 \\
\midrule
Total & 46 \\
\bottomrule
\end{tabular}
\end{table}

\subsection{Mutation Prompt Construction}
\label{app:mutation-prompt}

For each mutation request, the agent constructs a structured prompt containing four parts:

\begin{enumerate}
    \item the original MLIR program,
    \item the dialect tag,
    \item the selected mutation requirement,
    \item the dialect-specific inspection metadata.
\end{enumerate}

The prompt instructs the generation model to return only the mutated MLIR program. It also asks the model to preserve unrelated operations, maintain SSA consistency, respect dialect syntax, and apply only the requested mutation. This structure is important because small changes in MLIR can easily produce invalid SSA references, mismatched tensor types, or verifier failures.

\subsection{Validation and Regeneration}
\label{app:mutation-validation}

Because generated MLIR can be invalid, every candidate mutation is checked before it enters the benchmark dataset. The validation step verifies that the mutated program is syntactically well formed and compatible with the expected dialect-level constraints for the selected mutation type. If validation fails, the agent regenerates the mutation with the validation error included as feedback.

MLIRBench allows up to three generation attempts for each mutation. If all attempts fail, the candidate is discarded and the failure is recorded. Accepted mutations are forwarded to the Execution Sandbox Agent, where their final semantic labels are determined from compiler and execution behavior rather than from the intended mutation rule alone.

\subsection{Stored Mutation Records}
\label{app:mutation-records}

For each generated mutation, the pipeline stores both the accepted program pair and the metadata needed for later auditing. The stored fields include the program identifier, dialect, seed, mutation level, requirement identifier, requirement description, original MLIR, mutated MLIR, validation status, validation attempts, validation errors, failure category, and pipeline metadata.

\begin{table}[t]
\centering
\caption{Core fields stored for each Mutation Agent record.}
\label{tab:app_mutation_records}
\small
\begin{tabular}{p{0.32\columnwidth}p{0.58\columnwidth}}
\toprule
Field group & Stored fields \\
\midrule
Identity & Program ID, seed, dialect, level, requirement ID \\
Programs & Original MLIR and mutated MLIR \\
Validation & Status, attempts, errors, failure category \\
Requirement & Description and mutation constraints \\
Provenance & Pipeline metadata and validation history \\
\bottomrule
\end{tabular}
\end{table}

This record format makes it possible to trace a benchmark instance back to the mutation rule, generated MLIR, validation result, and any failed regeneration attempts.

\subsection{Common Mutation Failures}
\label{app:mutation-failures}

During benchmark construction, the most common mutation failures involved invalid SSA references, incorrect tensor shapes, malformed affine maps, unsupported operation variants, broken operand types, and edits that changed more of the program than the selected rule intended. These errors appeared more often in higher mutation levels, especially structural and failure-inducing mutations.

This behavior shows that generating plausible MLIR is easier than generating controlled compiler transformations. For this reason, validation and regeneration are necessary parts of the Mutation Agent rather than optional cleanup steps. The final benchmark includes only mutations that pass the required validation stage or are intentionally retained as failure-inducing cases for execution-grounded evaluation.

\section{Execution Agent: Implementation Details}
\label{app:execution-agent}
 
This appendix documents the compile flags, the upstream IREE crash that motivates
the dual-dialect input strategy, the full pipeline algorithm, and the status
breakdown behind the divergence labels reported in
Section 6.2.
\subsection{Compile Flags}
\label{app:compile-flags}
 
Both programs in a pair compile under identical IREE flags:
 
\begin{verbatim}
  --iree-llvmcpu-target-cpu=generic
  --iree-llvmcpu-loop-vectorization=false
  --iree-llvmcpu-slp-vectorization=false
  --iree-llvmcpu-loop-interleaving=false
  --iree-llvmcpu-loop-unrolling=false
\end{verbatim}
 
The four \texttt{=false} flags disable the LLVM transformations that reassociate
floating-point reductions: loop and SLP vectorization, loop interleaving, and
unrolling. The \texttt{target-cpu=\\generic} flag avoids architecture-specific code
paths such as AVX-512 that would make outputs depend on the host machine. Under
these flags both programs in a pair produce bit-stable output across repeated
runs, which is what licenses the strict equality comparison in
Section 3.3. We verify bit stability per pair rather than
across the full corpus; corpus-wide stability checking is left to future work.
 
\subsection{Compiler Crash on Embedded Torch Inputs}
\label{app:iree-bug}
 
Embedding a torch program's inputs as constants makes the compiler crash. A
program compiles cleanly when its input is a function argument, but once the input
is a \\ \texttt{torch.vtensor.literal} constant, the canonicalizer folds the
\texttt{torch.aten.lt.Scalar} comparison between that constant tensor and an
integer scalar, and torch-mlir's \\ \texttt{AtenLtScalarOp::fold} segfaults inside
its \\\texttt{comparisonScaleFolder} helper. The crash reproduces under all three
input-conversion modes the pipeline attempts (\texttt{auto}, \texttt{torch},
\texttt{tm\_tensor}). Because this comparison appears across our torch seeds, we
supply all torch inputs at runtime rather than embedding them.
 
\subsection{Pipeline Algorithm}
\label{app:execution-algorithm}
 
Algorithm~\ref{alg:sandbox} shows the success path.
 
\begin{algorithm}[t]
\caption{Execution Sandbox Agent. Compilation and execution use the
floating-point-deterministic flags of Appendix~\ref{app:compile-flags} and run in
isolated subprocesses. Failures are tagged \texttt{compile\_error},
\texttt{runtime\_error}, or, for Level~5 inputs, \texttt{expected\_invalid}.}
\label{alg:sandbox}
\begin{algorithmic}[1]
\Require pair $(A,B)$, dialect $d \in \{\textit{linalg}, \textit{torch}\}$
\Ensure record $(o_A, o_B, \delta, \tau, e)$
\State $A', I \gets \textsc{EmbedInputs}(A, d)$ \Comment{$\mathit{seed}=42$}
\State $B' \gets \textsc{EmbedInputs}(B, d, I)$ \Comment{same $I$}
\State $\hat{A} \gets \textsc{Compile}(A')$;\; $\hat{B} \gets \textsc{Compile}(B')$
\If{compile failure} \State \Return \texttt{compile\_error} \EndIf
\State $o_A \gets \textsc{Run}(\hat{A}, I)$;\; $o_B \gets \textsc{Run}(\hat{B}, I)$
\If{run failure} \State \Return \texttt{runtime\_error} \EndIf
\State $\delta \gets \lnot\,\textsc{ArrayEqual}(o_A, o_B)$ \Comment{true = diverge}
\State $\tau \gets \textsc{Benchmark}(\hat{A}, \hat{B}, I)$
\State $\phi \gets \textsc{StaticDiff}(A, B)$
\State $e \gets \textsc{Explain}(A, B, o_A, o_B, \delta, \phi)$ \Comment{$\phi$ blocks hallucinated SSA}
\State \Return $(o_A, o_B, \delta, \tau, e)$
\end{algorithmic}
\end{algorithm}
 
\subsection{Execution Status Breakdown}
\label{app:execution-status}

Table~\ref{tab:status} summarizes the execution status distribution across all released program pairs.
 
\begin{table}[t]
\centering
\caption{Execution status across the 1{,}768 released pairs. The 25
\texttt{runtime\_error} records are torch T2 mutations that altered the function
input signature, which kept the benchmarking subprocess from binding the seed's
input flags. Both programs executed in the ground-truth pipeline, so these
records are retained with a \texttt{runtime\_error\_category} field
(\texttt{input\_shape\_mismatch} or \texttt{input\_dtype\_mismatch}). The 3 \texttt{compile\_error} pairs are excluded from
the divergence table.}
\label{tab:status}
\small
\begin{tabular}{lrr}
\toprule
Status & Count & \% \\
\midrule
\texttt{success}           & 1{,}402 & 79.3 \\
\texttt{expected\_invalid} & 338     & 19.1 \\
\texttt{runtime\_error}    & 25      &  1.4 \\
\texttt{compile\_error}    & 3       &  0.2 \\
\midrule
Total                      & 1{,}768 & 100  \\
\bottomrule
\end{tabular}
\end{table}
 
\subsection{L1/T1 and L5 Anomalies}
\label{app:execution-anomalies}
 
Twenty L1/T1 pairs (1.1\% of the labeled set) diverge despite carrying a
syntactic level label. Manual inspection of the generated explanations attributes
these to two causes. Around twelve pairs contain constant-value changes or
attribute reorderings, which are semantic edits the upstream engine recorded as
syntactic. The remaining eight reorder independent reduction operations, which
changes the order of floating-point summation and produces a genuine difference
that strict comparison correctly reports. The second cause is the intended
behavior of the comparison design; the first reflects a limit of automated level
labeling at the source.
 
Seven linalg L5 pairs compiled and executed despite their intended-invalid label,
because IREE accepted the type-mismatched constructs through implicit coercion.
We keep these records with their observed status rather than relabeling them.


\section{LLM Reasoning Agent: Implementation Details}
\label{app:reasoning-agent}

This appendix provides implementation details for the LLM Reasoning Agent described in Section 3.4. The main paper presents the evaluation protocol and results; this appendix documents the prompt structure, visible and hidden artifacts, output schema, parsing and retry behavior, evaluated model roster, stored record format, and reproducibility controls used to make the evaluation auditable.

\subsection{Evaluation Scope}
\label{app:reasoning-scope}

The Reasoning Agent evaluates models under execution-hidden conditions. For each program pair, the model receives the two MLIR programs, the dialect tag, and deterministic input values when available, but it does not receive executed outputs, execution traces, lowered IR, ground-truth labels, or Judge Agent scores. The final evaluation uses 888 \texttt{linalg} pairs and 880 \texttt{torch} pairs. Each pair is evaluated independently by nine models, producing 15{,}912 structured reasoning records.

\begin{table}[h]
\centering
\caption{Reasoning Agent evaluation scale.}
\label{tab:app_reasoning_scale}
\small
\begin{tabular}{lrr}
\toprule
Dialect & Number of pairs & Total records \\
\midrule
\texttt{linalg} & 888 & 7{,}992 \\
\texttt{torch}  & 880 & 7{,}920 \\
\midrule
Total & 1{,}768 & 15{,}912 \\
\bottomrule
\end{tabular}
\end{table}

\subsection{Prompt Structure}
\label{app:reasoning-prompt}

The Reasoning Agent uses a predict-then-compare prompt. Rather than asking the model only to classify whether two programs are equivalent, the prompt first asks the model to reason about each program independently and then compare the predicted behaviors. This design separates per-program execution reasoning from pairwise semantic classification.

For each program, the model is instructed to inspect SSA value definitions and uses, tensor shapes and element types, affine indexing maps, iterator types, constants, operands, returned values, and dialect-specific operation semantics. The prompt is organized into four blocks:

\begin{enumerate}
\item \textbf{Task definition:} the model is told that it must determine the behavioral semantic relationship between Program $A$ and Program $B$.
\item \textbf{Isolation rule:} the model is reminded that it does not have access to executed outputs, lowered IR, execution traces, or the ground-truth label.
\item \textbf{Program context:} the prompt provides the dialect tag, deterministic inputs when available, Program $A$, and Program $B$.
\item \textbf{Output contract:} the model must return a single structured JSON object satisfying the required schema.
\end{enumerate}

When concrete inputs are available, they are included in the prompt as the inputs over which the model should reason. The executed outputs remain hidden. This allows input-grounded symbolic reasoning while preserving isolation between the Reasoning Agent and the Execution Sandbox Agent.

\subsection{Visible and Hidden Artifacts}
\label{app:reasoning-visible-hidden}

Table~\ref{tab:app_reasoning_visibility} summarizes which artifacts are visible to the Reasoning Agent and which artifacts are reserved for downstream evaluation. The key design choice is that the model sees the source-level MLIR pair and deterministic inputs, but all execution-derived evidence is withheld until scoring.

\begin{table*}[t]
\centering
\caption{Information visible to the Reasoning Agent versus information reserved for downstream evaluation.}
\label{tab:app_reasoning_visibility}
\small
\begin{tabular}{lcc}
\toprule
Artifact & Visible to Reasoning Agent & Used for Evaluation \\
\midrule
Program $A$ MLIR source & Yes & Yes \\
Program $B$ MLIR source & Yes & Yes \\
Dialect tag & Yes & Yes \\
Concrete input values & Yes, when available & Yes \\
Executed output of Program $A$ & No & Yes \\
Executed output of Program $B$ & No & Yes \\
Ground-truth semantic label & No & Yes \\
Lowered IR or execution trace & No & Optional \\
Judge score & No & Yes \\
\bottomrule
\end{tabular}
\end{table*}

\subsection{Output Schema}
\label{app:reasoning-schema}

Each model response must contain one JSON object. The required fields are listed in Table~\ref{tab:app_reasoning_schema}. The label space is intentionally restricted to \textsc{Equiv}, \textsc{Diverge}, and \textsc{Error}; an abstention label is not allowed. If a model is uncertain, it must still choose the best-supported label and lower its confidence.

\begin{table*}[t]
\centering
\caption{Required JSON output schema for each Reasoning Agent response.}
\label{tab:app_reasoning_schema}
\small
\begin{tabular}{lll}
\toprule
Field & Type & Description \\
\midrule
\texttt{predicted\_output\_A} & string/object & Predicted output or symbolic behavior for Program $A$ \\
\texttt{predicted\_output\_B} & string/object & Predicted output or symbolic behavior for Program $B$ \\
\texttt{label} & enum & One of \textsc{Equiv}, \textsc{Diverge}, or \textsc{Error} \\
\texttt{confidence} & float & Self-reported confidence in $[0,0.99]$ \\
\texttt{key\_differences} & list & Program-level semantic differences identified by the model \\
\texttt{witness} & string/object/null & Optional input or condition supporting divergence \\
\texttt{explanation} & string & Natural-language explanation of the predicted verdict \\
\bottomrule
\end{tabular}
\end{table*}

The prediction task can be written as:
\[
\hat{y} \in \{\textsc{Equiv}, \textsc{Diverge}, \textsc{Error}\}.
\]
The model must produce predicted behavior for each program before assigning $\hat{y}$. This ordering is intended to discourage surface-level comparison and make failures diagnosable: an incorrect final label can be separated from an incorrect output prediction.

\subsection{Parsing, Validation, and Retry Semantics}
\label{app:reasoning-retry}

Model outputs are validated against the required JSON schema before storage. Since language models sometimes produce malformed structured output, the parser performs lightweight recovery before declaring failure. The recovery procedure removes Markdown code fences, extracts the first balanced top-level JSON object, sanitizes unescaped control characters inside string fields, parses the JSON object, validates required fields and label values, and clamps confidence values greater than or equal to 1.0 to 0.99 while preserving the original value for audit.

Each response is assigned one of the status codes in Table~\ref{tab:app_reasoning_status}. If the first attempt does not produce a \texttt{SUCCESS} response, the agent retries once with a forced-verdict instruction. The retry instruction restates the required schema, disallows abstention, and requires the model to return a best-effort committed label. If the retry also fails, the record is finalized with a fallback \textsc{Error} label while preserving the underlying failure status for audit.

\begin{table}[t]
\centering
\caption{Reasoning Agent response status codes.}
\label{tab:app_reasoning_status}
\small
\setlength{\tabcolsep}{3pt}
\renewcommand{\arraystretch}{1.05}
\begin{tabular}{@{}p{0.30\columnwidth}p{0.62\columnwidth}@{}}
\toprule
Status & Meaning \\
\midrule
\texttt{SUCCESS} & Valid JSON satisfying the required schema \\
\texttt{PARSE\_ERROR} & No recoverable JSON object was found \\
\texttt{SCHEMA\_ERROR} & JSON parsed, but required fields were missing or malformed \\
\texttt{API\_ERROR} & Provider request failed \\
\texttt{MODEL\_FAILURE} & Model returned an explicit error payload \\
\bottomrule
\end{tabular}
\end{table}

\subsection{Model Roster}
\label{app:reasoning-models}

The evaluation uses a fixed roster of nine models: five open-source or open-weight systems accessed through OpenRouter-compatible endpoints, and four proprietary baselines accessed through native provider APIs when available. We select one representative per model family or provider to cover different training lineages rather than evaluating many closely related variants from the same lab.

\begin{table*}[t]
\centering
\caption{Model roster used in the Reasoning Agent evaluation.}
\label{tab:app_reasoning_models}
\small
\begin{tabular}{llll}
\toprule
Model & Provider/family & Open/closed & Role in comparison \\
\midrule
Qwen3-Next-80B-A3B-Instruct & Alibaba & Open & Code-oriented open baseline \\
Llama-4-Maverick & Meta & Open & General-purpose open baseline \\
GLM-5.1 & Zhipu AI & Open & Non-U.S.\ open model family \\
Kimi-K2.6 & Moonshot AI & Open & Long-context open model family \\
\texttt{gpt-oss-120b} & OpenAI OSS & Open & Large open-weight baseline \\
GPT-4.1-Mini & OpenAI & Closed & Cost-efficient proprietary baseline \\
Claude Sonnet 4.5 & Anthropic & Closed & Strong proprietary baseline \\
Claude Haiku 4.5 & Anthropic & Closed & Low-latency proprietary baseline \\
Gemini 3.1 Pro Preview & Google & Closed & Frontier reasoning model \\
\bottomrule
\end{tabular}
\end{table*}

All models are queried at temperature 0. Model identifiers are pinned through provider-specific model IDs or OpenRouter slugs at evaluation time. This reduces run-to-run variation and makes differences across systems more attributable to model behavior than to sampling randomness.

\subsection{Stored Reasoning Records}
\label{app:reasoning-records}

For each model and program pair, the Reasoning Agent stores a flat record containing the fields shown in Table~\ref{tab:app_reasoning_records}. These records intentionally separate model predictions from execution-derived ground truth. The Judge Agent later joins the reasoning record with the hidden execution record to compute hard-label accuracy, result-behavior accuracy, explanation quality, and aggregate score.

\begin{table*}[t]
\centering
\caption{Fields stored for each structured reasoning record.}
\label{tab:app_reasoning_records}
\small
\begin{tabular}{ll}
\toprule
Field group & Stored fields \\
\midrule
Pair metadata & \texttt{pair\_id}, \texttt{program\_id}, \texttt{mutation\_id}, dialect, mutation kind \\
Model metadata & model key, model identifier, model name, provider, model family \\
Prediction & predicted label, confidence, predicted outputs for Program $A$ and Program $B$ \\
Explanation & key differences, witness, natural-language explanation \\
Execution-hidden inputs & deterministic input values supplied to the prompt when available \\
Reliability metadata & status code, retry count, latency, raw-response head when parsing fails \\
Provenance & run timestamp and source result file \\
\bottomrule
\end{tabular}
\end{table*}

The flat record format is useful for auditing. A failed or suspicious score can be traced back to the exact model response, prompt-visible inputs, parsing status, retry count, and source pair identifier without rerunning the full benchmark.

\subsection{Reasoning Agent Procedure}
\label{app:reasoning-procedure}

The Reasoning Agent follows the same procedure for each model and program pair:

\begin{enumerate}
\item Build the prompt from Program $A$, Program $B$, the dialect tag, and deterministic inputs when available.
\item Invoke the evaluated model at temperature 0 using the pinned model identifier.
\item Parse and validate the response against the required JSON schema.
\item If parsing or schema validation fails, retry once with the forced-verdict instruction.
\item If the retry also fails, finalize the record with a fallback \textsc{Error} label while retaining the underlying failure status for audit.
\item Clamp confidence values greater than or equal to 1.0 to 0.99 while preserving the original value.
\item Store the structured reasoning record, including prediction fields, parsing status, retry count, latency, and provenance.
\end{enumerate}

This procedure avoids using the \texttt{algorithm} environment so that the appendix does not require additional LaTeX packages beyond the standard table and formatting packages already used in the paper.

\subsection{Reproducibility}
\label{app:reasoning-reproducibility}

The Reasoning Agent is designed as a reproducible evaluation component. For a fixed program pair, dialect, input set, prompt version, and pinned model version, the agent produces one structured record per model. The benchmark stores the prompt-visible inputs, model identifiers, parsed outputs, validation statuses, retry counts, latencies, and provenance metadata. This makes later auditing possible without rerunning all model calls and ensures that scoring remains separated from prediction generation.

\section{Judge Agent: Implementation Details}
\label{app:judge-agent}

This appendix documents the Judge Agent used to score MLIRBench model predictions. The main paper describes the high-level role of the Judge Agent; here we provide the evaluation object, prompt structure, scoring rubric, output schema, validation procedure, and stored-record format.

\subsection{Evaluation Object}
\label{app:judge-object}

The Judge Agent scores one evaluated-model prediction for one MLIR program pair. Each request contains the original program $A$, the mutated program $B$, dialect information, mutation metadata, the execution record produced by the Execution Sandbox Agent, and the structured prediction produced by the LLM Reasoning Agent.

The sandbox execution record is the only source of semantic ground truth. If either program fails compilation, verification, or execution, the normalized ground-truth label is \textsc{Error}. Otherwise, the output comparison determines whether the pair is \textsc{Equiv} or \textsc{Diverge}. The evaluated model's predicted label is normalized into the same three-class label space before scoring.

\begin{table}[H]
\centering
\caption{Inputs provided to each Judge Agent invocation.}
\label{tab:judge_inputs}
\small
\begin{tabular}{p{0.32\columnwidth}p{0.58\columnwidth}}
\toprule
Input group & Contents \\
\midrule
Program pair & Original MLIR and mutated MLIR \\
Dialect context & \texttt{torch} or \texttt{linalg} guidance \\
Mutation metadata & Mutation ID, level, rule description, and constraints \\
Execution record & Statuses, inputs, outputs, diagnostics, and divergence flag \\
Model prediction & Predicted label, outputs, confidence, and explanation \\
\bottomrule
\end{tabular}
\end{table}

\subsection{Judge Configuration}
\label{app:judge-config}

All benchmark predictions are scored with a single fixed judge model so that score differences reflect the evaluated model outputs rather than changes in the scorer. The judge is invoked in JSON-output mode and returns both compact scores and detailed evidence fields.

\begin{table}[t]
\centering
\caption{Judge Agent configuration.}
\label{tab:judge_config}
\small
\begin{tabular}{ll}
\toprule
Setting & Value \\
\midrule
Invocation unit & One prediction for one program pair \\
Default judge model & \texttt{gpt-5.4-mini} \\
Temperature & Not explicitly set \\
Response format & JSON object \\
Label space & \textsc{Equiv}, \textsc{Diverge}, \textsc{Error} \\
Returned fields & Scores, evidence, reasoning, verdict \\
\bottomrule
\end{tabular}
\end{table}

If the judge request fails or the returned JSON cannot be parsed, the pipeline constructs a conservative fallback record with zero result-behavior and explanation-quality scores. Hard-label accuracy remains deterministic because it is computed outside the judge model from the normalized prediction and execution-derived label.

\subsection{Prompt Blocks}
\label{app:judge-prompt-blocks}

The judge prompt is intentionally evidence-rich. It provides all source, execution, mutation, and prediction context in a single request so that the evaluated model's explanation does not become the primary source of facts.


\begin{table}[t]
\centering
\caption{Prompt blocks provided to the Judge Agent.}
\label{tab:judge_prompt_blocks}
\small
\setlength{\tabcolsep}{3pt}
\renewcommand{\arraystretch}{1.05}
\begin{tabular}{@{}p{0.24\columnwidth}p{0.36\columnwidth}p{0.30\columnwidth}@{}}
\toprule
Block & Examples & Purpose \\
\midrule
Program pair & Program $A$, Program $B$ & Source-level audit \\
Dialect knowledge & Torch/Linalg guidance & Dialect-aware reasoning \\
Mutation metadata & Level, rule, constraints & Scoring emphasis \\
Execution truth & Statuses, outputs, diagnostics & Ground truth \\
Prediction & Label, outputs, explanation & Object being scored \\
\bottomrule
\end{tabular}
\end{table}

For \texttt{linalg}, the prompt emphasizes destination-style semantics, indexing maps, iterator types, region bodies, \\ \texttt{linalg.yield}, result shapes, and element types. For \texttt{torch}, the prompt emphasizes ATen operator variants, operand order, dimension arguments, broadcasting, dtype annotations, reshape/view constraints, and SSA dataflow.

\subsection{Judge Prompt Template}
\label{app:judge-prompt-template}

The following is the condensed prompt template used by the Judge Agent. The actual prompt instantiates the template with full MLIR programs, execution records, mutation metadata, and model predictions.

\begin{lstlisting}[basicstyle=\ttfamily\scriptsize,breaklines=true]
System:
You are the MLIRBench Judge Agent, an expert MLIR compiler engineer.
Score the evaluated model strictly against the MLIR code and the
execution-derived ground truth. Do not reward fluent explanations unless
they are supported by the programs, dialect semantics, execution outputs,
or compiler diagnostics. Return JSON only.

User:
dialect: {linalg|torch|unknown}
pair_id: {pair_id}
mutation_id: {mutation_id}

--- Program A ---
{original MLIR}

--- Program B ---
{mutated MLIR}

--- Dialect Guidance ---
{dialect-specific MLIR guidance}

--- Mutation Metadata ---
mutation_level: {level}
mutation_rule: {rule_description}
mutation_constraints: {constraints}
structure_focus: {mutation-aware scoring profile}

--- Execution Ground Truth ---
inputs: {deterministic inputs}
status_A: {success|compile_error|runtime_error|expected_invalid}
status_B: {success|compile_error|runtime_error|expected_invalid}
output_A: {output or null}
output_B: {output or null}
outputs_diverge: {true|false|null}
diagnostics: {compiler/runtime diagnostics}
ground_truth_label: {EQUIV|DIVERGE|ERROR}
sandbox_explanation: {execution-grounded explanation}

--- Evaluated Model Prediction ---
predicted_label: {label}
predicted_output_A: {prediction}
predicted_output_B: {prediction}
confidence: {confidence}
key_differences: {differences}
witness: {witness}
explanation: {model explanation}

Task:
1. Build an independent semantic audit from the MLIR programs,
   dialect guidance, and execution evidence.
2. Compare the evaluated model prediction against the execution-derived
   ground truth.
3. Score result behavior and explanation quality.
4. Return one JSON object following the required schema.
\end{lstlisting}

The prompt orders the evidence before the evaluated model's explanation. This is intended to reduce anchoring on plausible but unsupported natural-language claims.

\subsection{Ground-Truth and Prediction Normalization}
\label{app:judge-normalization}

The Judge Agent normalizes execution and prediction labels before scoring. Execution labels are derived as follows:

\begin{itemize}
\item If either program has a non-success compilation, verification, or execution status, the ground-truth label is \textsc{Error}.
\item Otherwise, if the sandbox reports output divergence, the label is \textsc{Diverge}.
\item Otherwise, the label is \textsc{Equiv}.
\end{itemize}

Predicted labels are mapped into the same label space. Case differences, common spelling variants, and JSON string variants are normalized before comparison. Labels outside the supported set are treated as invalid and scored as non-matching.

\subsection{Scoring Dimensions}
\label{app:judge-scoring}

The final score combines three quantities: hard-label accuracy, result-behavior accuracy, and explanation quality.

\paragraph{Hard-label accuracy.}
Hard-label accuracy is computed deterministically:
\[
h = \mathbb{1}[\hat{y}=y],
\]
where $\hat{y}$ is the normalized predicted label and $y$ is the execution-derived label. This score does not depend on the judge model.

\paragraph{Result-behavior accuracy.}
Result-behavior accuracy $r \in [0,10]$ measures whether the model's predicted behavior matches execution. For successful executions, the judge scores output structure, exactness, and value precision. Output structure includes return arity, scalar/tensor category, rank, shape, and element type. Exactness rewards predictions that uniquely determine the observed outputs. Value precision gives partial credit for correctly predicted comparable scalar leaves.

For \textsc{Error} cases, result behavior is binary: $r=10$ if the prediction clearly recognizes compilation, verification, or runtime failure, and $r=0$ otherwise. The specific diagnostic quality is evaluated under explanation quality rather than result behavior.

\paragraph{Explanation quality.}
Explanation quality $q \in [0,4]$ follows a MASCoD-ICE-style rubric. The judge evaluates whether the explanation is useful, factually consistent, causal, evidence-specific, and grounded in the relevant MLIR dialect. Unsupported or hallucinated explanations receive low credit even when the final label is correct.

\begin{table}[t]
\centering
\caption{Explanation-quality anchors.}
\label{tab:judge_explanation_anchors}
\small
\begin{tabular}{cl}
\toprule
Score & Interpretation \\
\midrule
4 & Correct, causal, evidence-specific, and dialect-aware \\
3 & Mostly correct but missing one important detail \\
2 & Partially relevant but weak on causality or evidence \\
1 & Minimal relevant content with major omissions or errors \\
0 & Empty, generic, fabricated, or contradicted by evidence \\
\bottomrule
\end{tabular}
\end{table}

The aggregate score is:
\[
S = \frac{100}{3}
\left(
h + \frac{r}{10} + \frac{q}{4}
\right).
\]
This equal weighting separates coarse classification, concrete behavior prediction, and explanation quality.

\subsection{Mutation-Aware Result Rubric}
\label{app:judge-mutation-rubric}

For successful executions, the Judge Agent adjusts the structure component of the result-behavior score using mutation metadata. Shape- and structure-sensitive mutations receive more structure credit, while low-level syntactic edits emphasize exact values.

Let $p$ be the maximum number of structure points. The result-behavior score is:
\[
r = r_{\mathrm{struct}} + r_{\mathrm{exact}} + r_{\mathrm{value}},
\]
where $r_{\mathrm{struct}} \in [0,p]$, $r_{\mathrm{exact}} \in \{0,7-p\}$, and $r_{\mathrm{value}} \in [0,3]$.


\begin{table}[t]
\centering
\caption{Mutation-aware structure profiles used by the Judge Agent.}
\label{tab:judge_structure_profiles}
\small
\setlength{\tabcolsep}{2.5pt}
\renewcommand{\arraystretch}{1.05}
\begin{tabular}{@{}p{0.15\columnwidth}p{0.06\columnwidth}p{0.38\columnwidth}p{0.33\columnwidth}@{}}
\toprule
Profile & $p$ & Typical mutations & Emphasis \\
\midrule
Low & 1 & SSA renaming, formatting & Exact values \\
Default & 2 & General edits & Balanced scoring \\
Medium & 3 & Parameters, dtypes, operators & Mixed structure/value \\
High & 4 & Shapes, ranks, indexing maps & Structure-sensitive reasoning \\
Error & -- & Invalid MLIR or failed execution & Failure recognition \\
\bottomrule
\end{tabular}
\end{table}

The profile is inferred from the mutation rule text and mutation identifier. Rules mentioning shape, rank, broadcasting, indexing maps, or iteration space use the high-structure profile. Rules intended to introduce type, dependency, operator-constraint, or structural failures use the error profile.

\subsection{Deterministic Error-Case Post-Check}
\label{app:judge-error-postcheck}

The Judge Agent applies a deterministic post-check for \textsc{Error}-labeled records. It searches the evaluated model's normalized label, predicted outputs, and explanation for clear failure claims such as compilation failure, verification failure, runtime failure, invalid MLIR, unsupported operation, exception, or execution error. Negated phrases such as ``no error'' are handled so they do not trigger false failure recognition.

This post-check overwrites only the result-behavior score for error cases. It does not overwrite explanation quality. Thus, recognizing that a program fails is deterministic, while the quality of the diagnostic explanation remains judged from the full MLIR and execution evidence.

\subsection{Required Judge Output Schema}
\label{app:judge-output-schema}

The judge returns a single JSON object. The parser strips optional Markdown code fences, extracts the outer JSON object if necessary, parses the response, clamps score ranges, and stores both compact and detailed records.

\begin{lstlisting}[basicstyle=\ttfamily\scriptsize,breaklines=true]
{
  "scores": {
    "result_accuracy": <number from 0 to 10>,
    "explanation_rating": <integer from 0 to 4>
  },
  "evidence": {
    "correct_claims": ["..."],
    "incorrect_claims": ["..."],
    "unverifiable_claims": ["..."],
    "semantic_audit": ["..."],
    "result_accuracy_components": {
      "mode": "success" | "error",
      "shape_structure_points": <number or null>,
      "shape_structure_max": <number or null>,
      "exactness_points": <number or null>,
      "exactness_max": <number or null>,
      "value_precision_points": <number or null>,
      "error_prediction_points": <number or null>,
      "calculation": "..."
    },
    "explanation_subcriteria": {
      "usefulness": "...",
      "factual_consistency": "...",
      "semantic_causality": "...",
      "evidence_specificity": "...",
      "dialect_grounding": "..."
    }
  },
  "reasoning": "...",
  "verdict": "CORRECT" | "PARTIALLY_CORRECT" | "INCORRECT"
}
\end{lstlisting}

\subsection{Verdict Assignment}
\label{app:judge-verdicts}

After parsing the judge output, MLIRBench computes the aggregate score and assigns a categorical verdict. A record is marked \textsc{Correct} when the predicted label matches the execution-derived label, result-behavior accuracy is at least 8, and explanation quality is at least 3. A record is marked \textsc{Partially Correct} when the aggregate score is at least 50 or when at least one major component shows meaningful evidence of correctness. All remaining records are marked \textsc{Incorrect}.

These verdicts are used only for qualitative analysis and error inspection. The primary quantitative metric is the aggregate score and its three subscores.

\subsection{Stored Judge Records}
\label{app:judge-records}

The Judge Agent stores both compact and detailed outputs. Compact records are used for aggregate reporting, while detailed records preserve the evidence needed for auditing individual scores.

\begin{table*}[t]
\centering
\caption{Fields stored in Judge Agent outputs.}
\label{tab:judge_record_fields}
\small
\begin{tabular}{ll}
\toprule
Field group & Stored fields \\
\midrule
Identity & Evaluated model, pair ID, program ID, mutation ID, dialect \\
Labels & Ground-truth label, predicted label, label match \\
Subscores & Hard-label accuracy, result accuracy, explanation rating \\
Weighted scores & Weighted label, weighted result behavior, weighted explanation \\
Aggregate & Total score, categorical verdict \\
Evidence & Semantic audit, correct claims, incorrect claims, unverifiable claims \\
Execution details & Inputs, statuses, outputs, divergence details, sandbox explanation \\
Prediction details & Predicted outputs, confidence, key differences, witness, explanation \\
Reliability metadata & Judge parse status, fallback status, retry or error information \\
\bottomrule
\end{tabular}
\end{table*}

\subsection{Validation Checks}
\label{app:judge-validation}

Before aggregate results are produced, the Judge Agent outputs are validated with the following checks:

\begin{enumerate}
\item Every prediction joins to exactly one execution record and one MLIR program pair.
\item Every judge response parses as JSON or is replaced by a conservative fallback record.
\item Required score, evidence, reasoning, and verdict fields are present.
\item Result-accuracy scores lie in $[0,10]$ and explanation ratings lie in $[0,4]$.
\item \textsc{Error} examples pass the deterministic failure-recognition post-check.
\item The final aggregate score is recomputed from stored subscores rather than trusted from the raw judge response.
\end{enumerate}

These checks make the judge outputs auditable and prevent malformed responses from silently affecting aggregate metrics.

\subsection{Judge Agent Algorithm}
\label{app:judge-algorithm}

Algorithm~\ref{alg:judge_agent} summarizes the Judge Agent scoring procedure.

\begin{algorithm}[H]
\caption{Judge Agent Scoring}
\label{alg:judge_agent}
\begin{algorithmic}[1]
\Require MLIR programs $A,B$; execution record $G$; prediction $P$
\Ensure Score record with subscores, aggregate score, verdict, and evidence
\State Extract program pair, dialect, mutation metadata, $G$, and $P$
\State Derive execution-grounded label $y$ from $G$
\State Normalize predicted label $\hat{y}$ from $P$
\State Compute hard-label score $h \gets \mathbb{1}[\hat{y}=y]$
\State Select dialect guidance and mutation-aware rubric profile
\State Build judge prompt from $A$, $B$, metadata, $G$, and $P$
\State Invoke fixed judge model with JSON response format
\If{response is missing or malformed}
    \State Construct conservative fallback judge result
\EndIf
\State Parse result-behavior score $r$, explanation score $q$, evidence, and verdict
\State Clamp $r$ to $[0,10]$ and $q$ to $[0,4]$
\If{$y$ is \textsc{Error}}
    \State Apply deterministic failure-recognition post-check to $r$
\EndIf
\State Recompute aggregate score $S$
\State Assign final categorical verdict
\State Store compact score row and detailed evidence record
\end{algorithmic}
\end{algorithm}

\subsection{Judge Reliability and Score Sensitivity}
\label{app:judge-failure-modes}
\label{app:judge-reliability}

We evaluated the sensitivity of model-level scores to repeated judging, judge selection, and scoring configuration. The study uses 900 frozen responses from nine candidate models on the same 100 program pairs. We completed three primary-judge runs, one alternative-judge run, a run without the generated reference narrative, and a run with fixed structure/exactness/value maxima of 2/5/3. The primary judge is GPT-5.4 Mini; the alternative is GPT-4.1. Both are from OpenAI.

The table reports mean aggregate scores on the 0--100 scale. "Original" denotes the historical grades restricted to these identical responses, rather than the full-benchmark scores. This matched comparison separates judge variation from differences in sample composition.

\begin{table*}[t]
\centering
\caption{Model-level aggregate scores (0--100) on identical responses to the same 100 program pairs. Original denotes historical grades on this sample; R1--R3 denote fresh primary-judge repeats. Alt. changes the judge, No ref. removes the generated reference narrative, and Fixed uses structure/exactness/value maxima of 2/5/3.}
\label{tab:judge-model-comparison}
\small
\setlength{\tabcolsep}{4pt}
\begin{tabular}{lrrrrrrr}
\toprule
Candidate model & Original & R1 & R2 & R3 & Alt. & No ref. & Fixed \\
\midrule
Gemini 3.1 Pro Preview & 62.51 & 61.77 & 62.14 & 61.89 & 67.55 & 62.67 & 61.75 \\
GPT-OSS 120B & 53.15 & 54.30 & 52.70 & 53.50 & 60.31 & 54.09 & 52.71 \\
Claude Sonnet 4.5 & 52.51 & 52.88 & 52.57 & 51.59 & 58.55 & 53.03 & 51.45 \\
GLM 5.1 & 50.53 & 49.77 & 50.74 & 49.59 & 55.87 & 50.16 & 49.07 \\
Kimi K2.6 & 49.20 & 49.34 & 49.09 & 48.93 & 56.02 & 48.78 & 46.94 \\
Claude Haiku 4.5 & 46.98 & 47.80 & 47.27 & 47.09 & 53.45 & 47.70 & 45.91 \\
GPT-4.1 Mini & 47.24 & 46.90 & 47.40 & 47.04 & 51.63 & 47.80 & 45.49 \\
Qwen3 Next 80B & 43.19 & 43.18 & 43.45 & 43.93 & 47.84 & 44.20 & 42.20 \\
Llama 4 & 38.18 & 39.27 & 38.55 & 38.22 & 40.48 & 39.66 & 37.07 \\
\bottomrule
\end{tabular}
\end{table*}

Across all nine candidate models, each fresh primary-judge run remains within 1.15 aggregate points of the original matched-sample score. Relative to the first repeat, removing the reference narrative changes model means by at most 1.02 points, while fixing the rubric lowers them by 0.01--2.41 points. These results support repeatability of model-level aggregates under the primary judge and limited aggregate sensitivity to reference removal on this sample.

The alternative judge assigns higher model-level scores, with increases of 1.21--6.67 points relative to the first repeat, indicating greater scoring leniency on average. Nevertheless, the three highest-ranked models retain their order across all six conditions, with only adjacent rank changes elsewhere. Thus, broad comparative conclusions are more stable than absolute score values.

This evidence concerns aggregate stability, not correctness of every individual judgment. The aggregate includes an unchanged hard-label component, and averaging can conceal individual disagreements. The alternative judge also produced internally inconsistent result components in 111 of 738 successful-execution judgments; these outputs remain included in the comparison. Human-expert alignment remains a separate validation step.

\section{Additional Results and Analysis}
\label{app:additional-results}

This appendix provides additional analysis for the benchmark results reported in the main paper. The main results focus on aggregate model performance, while this section gives more detail about how performance varies across mutation levels, dialects, and scoring dimensions.

\subsection{Evaluation Scale}

The final reasoning evaluation contains 1{,}768 MLIR program pairs: 888 from the \texttt{linalg} dialect and 880 from the \texttt{torch} dialect. Each pair is evaluated with nine language models, producing 15{,}912 structured reasoning records. Each record stores the model prediction, predicted outputs for both programs, confidence, explanation, parsing status, retry count, and the final Judge Agent scores.


\subsection{Aggregate Scoring Trends}

Across all models and program pairs, the mean aggregate score is 57.21. The mean hard-label accuracy is 74.59\%, while the mean result-behavior score is only 3.63 out of 10 and the mean explanation-quality score is 2.43 out of 4. This shows that models are often able to select the correct coarse semantic label, but they are less reliable at predicting the concrete behavior that justifies the label.

\subsection{Mutation-Level Behavior}

Model performance varies across mutation levels and model
families. Syntactic, parameter, operator, and structural
mutations test different aspects of program reasoning, while
failure-inducing mutations test compiler legality, type and
shape constraints, and verifier behavior. Failure-inducing
mutations challenge several models, but others achieve their
highest scores on this category. Mutation levels therefore
do not define a uniform difficulty ordering.

Strong label accuracy does not necessarily imply accurate
output prediction or failure diagnosis. A model may correctly
classify a pair as \textsc{EQUIV}, \textsc{DIVERGE}, or
\textsc{ERROR} while misidentifying output values, tensor
shapes, or the cause of failure. MLIRBench distinguishes these
abilities through separate measures of hard-label accuracy,
result behavior, and explanation quality.

\section{Downstream Task Details}
\label{app:downstream-tasks}

This appendix provides the prompt construction, data-splitting, and evaluation
details for the two downstream tasks in Section~7 that could not be included
in the main paper. 

\subsection{Fine-Tuning Configuration}
\label{app:dt-finetuning}

Both downstream tasks use
\path{Qwen2.5-Coder-14B-Instruct} as the base model and LoRA for supervised
fine-tuning. Table~\ref{tab:app_dt_training} summarizes the final training
configuration.

\begin{table}[t]
\centering
\caption{Final supervised fine-tuning configuration for the downstream tasks.}
\label{tab:app_dt_training}
\footnotesize
\setlength{\tabcolsep}{3pt}
\begin{tabularx}{\columnwidth}{@{}l >{\raggedright\arraybackslash}X >{\raggedright\arraybackslash}X@{}}
\toprule
Setting & Task 1 & Task 2 \\
\midrule
Base model
    & Qwen2.5-Coder-14B
    & Qwen2.5-Coder-14B \\
LoRA rank
    & 16
    & 64 \\
LoRA alpha
    & 32
    & 128 \\
LoRA dropout
    & 0.0
    & 0.0 \\
LoRA targets
    & q/v projections
    & attention + MLP projections \\
Learning rate
    & $1\times10^{-4}$
    & $1\times10^{-4}$ \\
Epochs
    & 3
    & 6 \\
Per-device batch size
    & 1
    & 1 \\
Gradient accumulation
    & 4
    & 1 \\
Maximum sequence length
    & 32{,}768
    & 32{,}768 \\
Precision
    & BF16
    & BF16 \\
Warmup steps
    & 20
    & 20 \\
Gradient checkpointing
    & Yes
    & Yes \\
\bottomrule
\end{tabularx}
\end{table}


\subsection{Task 1: Optimization-Benefit Prediction}
\label{app:dt1}

\subsubsection{Target Construction}

For \texttt{linalg}, we apply
\path{linalg-fuse-elementwise-ops} and measure the reduction in the number of
Linalg operations. We use the following benefit classes:

\begin{itemize}
    \item \textsc{Low}: 0--6 operations removed,
    \item \textsc{Medium}: 7--11 operations removed,
    \item \textsc{High}: 12 or more operations removed.
\end{itemize}

For \texttt{torch}, we apply
\path{torch-simplification-pipeline} and measure the change in total operation
count:

\begin{itemize}
    \item \textsc{Low}: no operation reduction or an increase in operation count,
    \item \textsc{Medium}: 1--2 operations removed,
    \item \textsc{High}: 3 or more operations removed.
\end{itemize}

The transformed IR, post-pass operation count, and compiler-derived benefit
class are used only to construct the target. They are not provided to the
model.

\subsubsection{Family-Disjoint Cross-Validation}

Programs are grouped by their underlying source-program family. Variants from
different mutation levels or rules that originate from the same source program
are kept in the same fold. For example, related variants are reduced to a
family identifier such as \texttt{seed2\_prog14}.

We use five family-disjoint folds, with each family appearing in a held-out
fold exactly once. Training and held-out family sets therefore have zero
overlap. Final cross-validation metrics are computed from the combined
held-out predictions across all five folds.

For the Torch SFT experiment, the original training distribution is
imbalanced. We therefore balance only the training partition by resampling
the minority classes. Validation and held-out folds are never resampled and
retain their original class distributions.

\subsubsection{Prompt Format}

For all evaluation settings, the model receives only information available
before the compiler transformation is applied. The prompt contains the
compiler transformation being predicted, the corresponding class boundaries,
static structural features extracted from the original program, and the
complete original MLIR program.

A representative Task~1 prompt is shown below.

\begin{lstlisting}[
basicstyle=\ttfamily\scriptsize,
breaklines=true,
columns=fullflexible
]
Predict the structural optimization benefit of applying
the specified compiler transformation to the MLIR program
below.

Benefit is defined using the compiler-observed reduction
in operations.

A = LOW
B = MEDIUM
C = HIGH

Use only the static structure of the original program.
Do not assume that the transformation is always beneficial.
Do not simulate or invent post-pass measurements.

Static structural features:
{static_features}

MLIR program:
{mlir_program}

Return exactly one token: A, B, or C.
\end{lstlisting}

The transformation name and benefit thresholds are changed according to the
dialect. For example, the Torch prompt specifies
\path{torch-simplification-pipeline} together with the corresponding
operation-count thresholds.

The static structural features are supplemental to the complete MLIR program
and contain counts of relevant dialect operations and structural tokens.

The transformed program, compiler-observed operation reduction, and target
class are never included in the input prompt.

\subsubsection{Zero-Shot and Three-Shot Evaluation}

The zero-shot condition contains only the task description and target program.

For three-shot evaluation, one demonstration is selected from each of
\textsc{Low}, \textsc{Medium}, and \textsc{High}. All demonstrations are
selected only from the training families of the current fold.

The demonstration format is identical to the target format and ends with the
corresponding single-token answer:

\begin{verbatim}
A
B
C
\end{verbatim}

No demonstration is selected from the held-out family.

\subsubsection{Supervised Fine-Tuning}

For SFT, the model receives the same information available in the prompting
experiments. The target is the corresponding one-token benefit class.

No transformed IR, post-pass operation count, compiler output, or measured
benefit is included in the training input. This keeps the prediction task
consistent across zero-shot, few-shot, and fine-tuned settings.

\subsubsection{Evaluation Details}

Macro-F1 is used as the main metric because the benefit-class distributions
are not balanced. We additionally compute:

\begin{itemize}
    \item accuracy,
    \item balanced accuracy,
    \item per-class F1,
    \item confusion matrices.
\end{itemize}

A prediction is valid only if it resolves to one of the three allowed labels.
Outputs outside the allowed label set are counted as incorrect rather than
removed from the evaluation.

\begin{figure}[t]
    \centering
    \includegraphics[width=\linewidth]
    {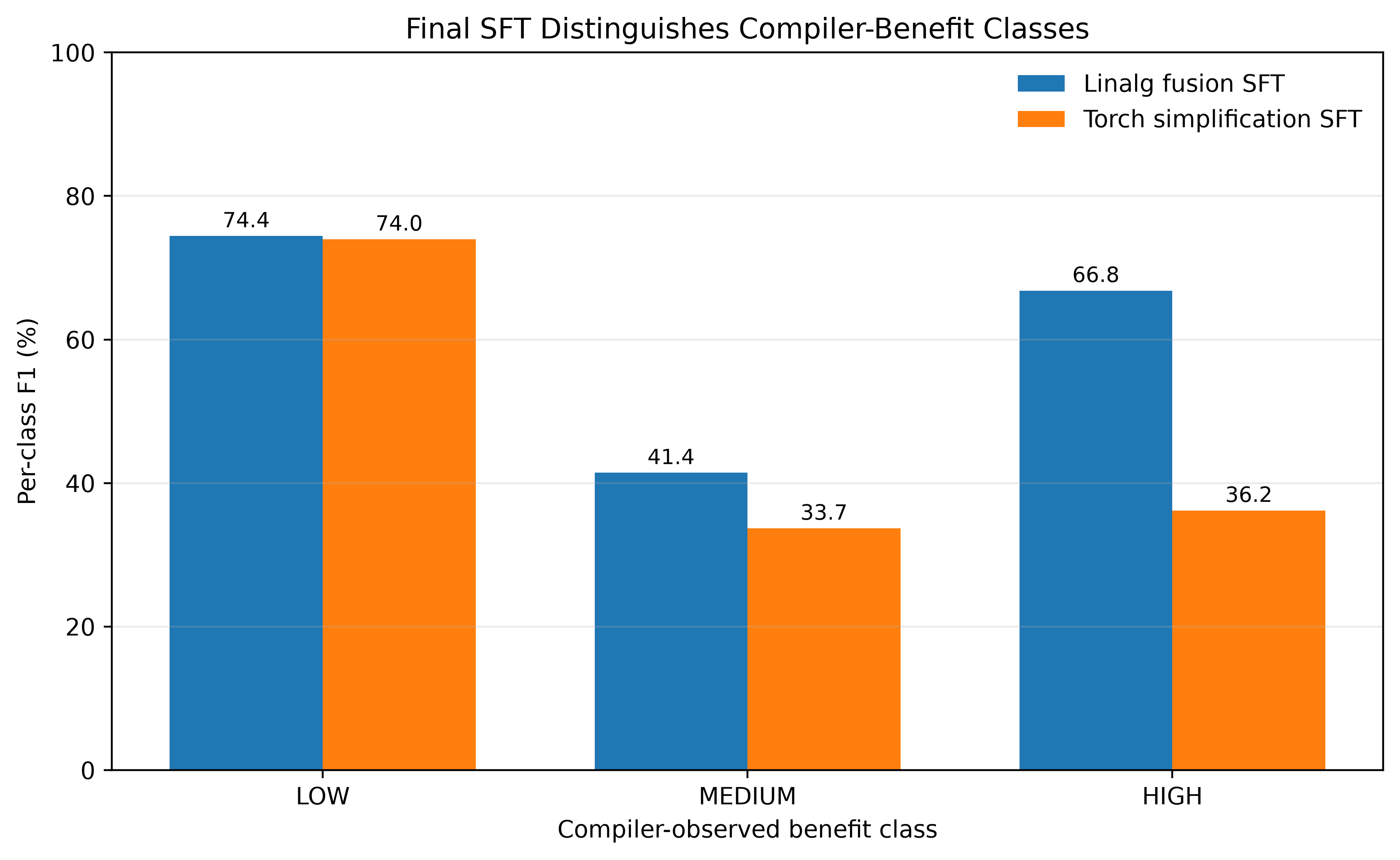}
    \caption{Per-class F1 for the final Task~1 SFT models. The figure shows
    performance separately for \textsc{Low}, \textsc{Medium}, and
    \textsc{High} compiler-observed optimization benefit.}
    \label{fig:app_dt1_classes}
\end{figure}

Figure~\ref{fig:app_dt1_classes} provides an additional per-class view of the
final SFT models that is not included in the main paper.

\subsubsection{RISC-V Backend Comparison}

The additional RISC-V analysis is performed only for the Linalg fusion
experiment.

For each compiler-valid program, the original and fusion-transformed versions
are lowered through the same RISC-V backend configuration. Their generated
static instruction streams are then compared.

The backend outcome is defined as:

\begin{itemize}
    \item \textit{improved}: the transformed program produces fewer static
    instructions,
    \item \textit{tied}: both versions produce the same number of static
    instructions,
    \item \textit{worsened}: the transformed program produces more static
    instructions.
\end{itemize}

This analysis uses generated static instruction counts only. It is not used as
a measurement of runtime or cycle-level performance.


\subsection{Task 2: Semantic Supervision Transfer}
\label{app:dt2}

\subsubsection{Family Construction and Data Isolation}

Task~2 uses MLIRBench program pairs as the source dataset. Program variants are
grouped by their underlying source-program family rather than by individual
mutation.

For example,

\begin{verbatim}
seed2_linalg_prog14_L1_R4
\end{verbatim}

and other mutations derived from the same source program are assigned to the
same family. Corresponding variants across the two dialects are also kept
within the same source family.

Train, validation, and test partitions are therefore disjoint at the family
level:

\begin{verbatim}
train families intersect validation families = empty
train families intersect test families       = empty
validation families intersect test families  = empty
\end{verbatim}

This prevents the model from training on one mutation of a source program and
then being evaluated on another related mutation from the same source family.

The split is also constructed to retain examples from both
\texttt{linalg} and \texttt{torch} and from all three semantic labels.

\subsubsection{Zero-Shot Prompt}

Zero-shot evaluation provides:

\begin{enumerate}
    \item the task instruction,
    \item the MLIR dialect,
    \item the original MLIR program,
    \item the mutated MLIR program.
\end{enumerate}

No execution result, benchmark ground-truth label, reference explanation,
Mutation Agent reasoning, or Judge Agent score is included.

The prompt follows this format:

\begin{lstlisting}[
basicstyle=\ttfamily\scriptsize,
breaklines=true,
columns=fullflexible
]
You are an MLIR compiler reasoning expert.

Given an original MLIR program and a mutated MLIR
program, predict their semantic relationship.

Allowed labels:
EQUIV
DIVERGE
ERROR

MLIR dialect:
{dialect}

Original MLIR:
{original_mlir}

Mutated MLIR:
{mutated_mlir}

Return exactly one label:
EQUIV, DIVERGE, or ERROR.
\end{lstlisting}

\subsubsection{Balanced Six-Shot Prompt}

The six-shot condition uses the same target prompt, but six labeled
demonstrations are added before the target program pair.

The demonstrations are balanced across both dialect and semantic class:

\begin{verbatim}
linalg / EQUIV
linalg / DIVERGE
linalg / ERROR

torch  / EQUIV
torch  / DIVERGE
torch  / ERROR
\end{verbatim}

All demonstrations are selected only from the training partition. No
validation or test family is used as an in-context demonstration.

The demonstrations use the same original-program, mutated-program, and
single-label format as the target example.

The six-shot experiment is evaluated using demonstration seeds 42 and 123.
The reported six-shot result is the aggregate across these two selections
rather than relying on a single demonstration set.

\subsubsection{Supervised Fine-Tuning}

The SFT condition uses the same program-pair representation as the prompting
baselines. The input contains:

\begin{itemize}
    \item MLIR dialect,
    \item original MLIR,
    \item mutated MLIR.
\end{itemize}

The supervised target is exactly one of:

\begin{verbatim}
EQUIV
DIVERGE
ERROR
\end{verbatim}

Execution outputs, execution traces, reference explanations, and Judge Agent
scores are not exposed to the downstream model during training or evaluation.

This keeps Task~2 focused on whether the semantic relationship encoded by the
MLIRBench label can be learned from the program pair itself.

\begin{figure}[!t]
    \centering
    \includegraphics[
        width=0.95\columnwidth,
        keepaspectratio
    ]{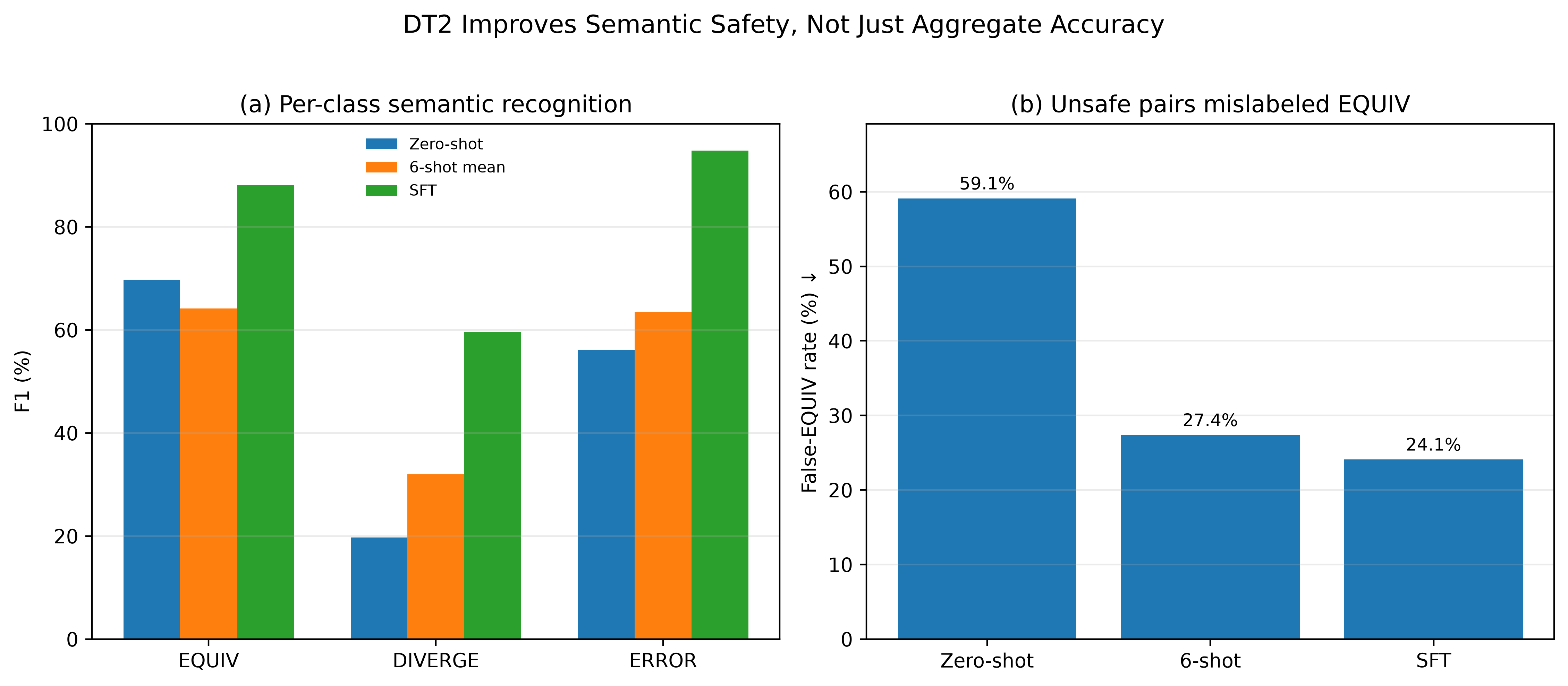}
    \caption{Additional Task~2 analysis. Left: per-class F1 for
    \textsc{EQUIV}, \textsc{DIVERGE}, and \textsc{ERROR}. Right:
    false-equivalence rate, measuring unsafe \textsc{DIVERGE} or
    \textsc{ERROR} pairs predicted as \textsc{EQUIV}.}
    \label{fig:app_dt2_safety}
\end{figure}

\subsubsection{Evaluation Details}

All evaluation settings use the same held-out source families.

We report:

\begin{itemize}
    \item accuracy,
    \item macro-F1,
    \item per-class F1,
    \item per-dialect macro-F1.
\end{itemize}

We additionally compute the false-equivalence rate. This captures cases where
an unsafe pair, i.e., a pair labeled \textsc{Diverge} or \textsc{Error}, is
incorrectly predicted as \textsc{Equiv}.

For ground-truth label $y$ and prediction $\hat{y}$, we define:

\[
\mathrm{FER}
=
\frac{
\#\left\{
y \in
\{\textsc{Diverge},\textsc{Error}\}
\land
\hat{y}=\textsc{Equiv}
\right\}
}{
\#\left\{
y \in
\{\textsc{Diverge},\textsc{Error}\}
\right\}
}.
\]

The false-equivalence rate is computed only from held-out predictions.

Figure~\ref{fig:app_dt2_safety} provides additional class-level and
false-equivalence analysis that is not included in the main paper.

\end{document}